\documentclass[aps, prb, twocolumn, superscriptaddress, amsmath,  tightenlines, longbibliography]{revtex4-1}

\usepackage{dcolumn}
\usepackage{graphicx}
\usepackage{mathrsfs}
\usepackage{subfigure}
\usepackage{booktabs}
\usepackage{amsmath}
\usepackage{physics}
\usepackage{dsfont}
\usepackage{amstext}
\usepackage{amssymb}
\usepackage{amsbsy}
\usepackage{bbm}
\usepackage{amsthm}
\usepackage{graphicx}
\usepackage{color}
\usepackage[colorlinks,citecolor=blue]{hyperref}

\usepackage{url}
\usepackage[colorlinks]{hyperref}
\hypersetup{%
	plainpages=true,
	breaklinks=true,       
	hypertexnames=false,  
	pageanchor=true,
	colorlinks=true,
	linkcolor={blue},
	citecolor={red},
	urlcolor={blue},
	anchorcolor={black}
}

 \makeatletter

\newcommand{\Rmnum}[1]{\expandafter\@slowromancap\romannumeral #1@}
\makeatother

\begin{document}
	
\title{Topological quantum transition driven by charge-phonon coupling in higher-order topological insulators}
\author{Congwei Lu}
\affiliation{Department of Physics, Applied Optics Beijing Area Major Laboratory, Beijing Normal University, Beijing 100875, China}
\author{Mei Zhang}
\affiliation{Department of Physics, Applied Optics Beijing Area Major Laboratory, Beijing Normal University, Beijing 100875, China}
\author{Haibo Wang}
\affiliation{Department of Physics, Applied Optics Beijing Area Major Laboratory, Beijing Normal University, Beijing 100875, China}
\author{Qing Ai}
\affiliation{Department of Physics, Applied Optics Beijing Area Major Laboratory, Beijing Normal University, Beijing 100875, China}
\author{Tao Liu}
\email[E-mail: ]{liutao0716@scut.edu.cn}
\affiliation{School of Physics and Optoelectronics, South China University of Technology,  Guangzhou 510640, China}

\date{{\small \today}}


\begin{abstract}
We investigate a second-order topological quantum transition of a modified Kane-Mele model driven by electron-phonon interaction. The results show that the system parameters of the bare modified Kane-Mele model are renormalized by the electron-phonon interaction. Starting from the second-order topological phase for the bare model, the increasing electron-phonon coupling strength can drive the second-order topological insulator into a semimetal phase. Such a second-order topological phase transition is characterized by the band-gap closing, discontinuity of averaged ferminoic number and topological invariant.
\end{abstract}
	
\maketitle 
	
\section{\label{sec:introduction}Introduction}
The past years have witnessed considerable interests and rapid developments in studying higher-order topological insulators (HOTIs) \cite{PhysRevLett.110.046404,Benalcazar61,PhysRevB.96.245115,PhysRevLett.119.246401,PhysRevLett.119.246402,PhysRevB.97.241405,peterson2018quantized,serra2018observation,arXiv:1801.10053,PhysRevLett.120.026801,schindler2018higher,Zhang2019,Ni2018,xue2019acoustic,arXiv:1801.10050,arXiv:1802.02585,PhysRevLett.123.216803,PhysRevResearch.2.033029,PhysRevLett.124.036803,PhysRevB.101.241104,PhysRevLett.124.063901,PhysRevB.103.L201115,arXiv:2202.12151,PhysRevLett.126.066401,PhysRevB.92.085126,PhysRevB.104.134508}. The hallmark of HOTIs is the existence of topologically-protected boundary states with their dimension at least two lower than the bulk states. Such unconventional boundary states have been experimentally observed in a variety of platforms, including electrical circuits\cite{peterson2018quantized,imhof2018topolectrical,PhysRevB.99.020304,PhysRevLett.126.146802}, acoustic\cite{xue2020observation,Ni2018,xue2019acoustic,Gao_2021} and photonic waveguids\cite{Mittal2019,el2019corner,Li2019}, phononic metamaterials\cite{serra2018observation}, and solid-state materials\cite{arXiv:1802.02585}. Furthermore, the unconventional bulk-boundary correspondence in HOTIs, with the interplay of disorder\cite{PhysRevResearch.2.033521,PhysRevLett.125.166801,PhysRevB.103.085408}, quasicrystal\cite{PhysRevLett.124.036803,PhysRevResearch.2.033071,PhysRevB.102.241102}  and amorphous\cite{PhysRevResearch.2.012067,PhysRevLett.126.206404} structures, many-body interactions\cite{PhysRevB.102.045110,PhysRevLett.123.196402,PhysRevLett.127.176601}, non-Hermiticity\cite{PhysRevLett.122.076801,PhysRevLett.123.073601,PhysRevB.103.224203} or periodic driving\cite{PhysRevLett.124.057001,PhysRevLett.124.216601,PhysRevB.102.094305,PhysRevB.101.235403,PhysRevB.105.115418,PhysRevB.103.115308,PhysRevB.105.155406,PhysRevB.105.115418}, has led to many intriguing features uncovered in conventional topological insulators.

Up to now, much effort has been devoted to understanding the effects of disorders\cite{PhysRevResearch.2.033521,PhysRevLett.125.166801,PhysRevB.103.085408} and electron-electron interactions\cite{PhysRevB.102.045110,PhysRevLett.123.196402,PhysRevLett.127.176601} on the higher-order topological properties. In contrast, the inevitable roles played by the electron-phonon interaction  in solid-state materials remain mostly unexplored in HOTIs. Therefore, it is of interest to address this issue. In conventional first-order topological phases, the electron-phonon interaction has been proven to modify the topological properties of an electronic structure, and induce novel topological phase transitions and robust edge states\cite{PhysRevLett.110.046402,2017Type,PhysRevLett.121.090402,PhysRevLett.117.246401,PhysRevLett.123.046401,PhysRevB.98.035423,PhysRevResearch.2.043431,PhysRevLett.128.066801}. One may wonder how the an electron-phonon interaction modifies the topological properties in HOTIs, and whether a higher-order topological  phase occurs by tuning the phononic degrees of freedom.

In this paper, we aim to reveal the role played by the electron-phonon coupling of solid-state materials in determining the higher-order topological phase transitions. To be specific, we  investigate  an Holstein model by introducing the  electron-phonon interaction into the modified  Kane-Mele model \cite{PhysRevLett.124.166804}. The renormalization of the system's parameters due to the electron-phonon coupling is demonstrated by using the Lange-Firsov approach \cite{PhysRevB.66.075129} in high-frequency limit of the optical phonon mode. Then we employ the cluster perturbation theory  to calculate the one-electron spectral function defined via the system's Green's function. We compute the renormalized parameters, band gaps, and energy bands as the electron-phonon coupling strength varies. The electron-phonon interaction modifies the system's parameters, and therefore the band gap closes at the critical electron-phonon coupling strength. These indicate that a topological phase transition occurs. Such a phase transition is further verified by analyzing the fermionic number discontinuity at the critical electron-phonon coupling strength. By calculating the topological invariant, we verify a second-order topological phase transition driven by the electron-phonon coupling.

The rest of the paper is structured as follows. In Sec.~\ref{sec:the model}, we review the bare modified Kane-Mele model and its topological phase regimes. In Sec.~\ref{sec:Charge-phonon coupling}, we introduce the Lang-Firsov approach to predict the renormalized parameters, and utilize the cluster perturbation theory to calculate the spectral function and second-order topological phase transitions. The numerical results are presented in Sec.~\ref{sec:Results}. Finally, we conclude the work in Sec.~\ref{sec:Conclusion}.

\begin{figure}[!tb]	
	\centering
	\includegraphics[width=7.8cm]{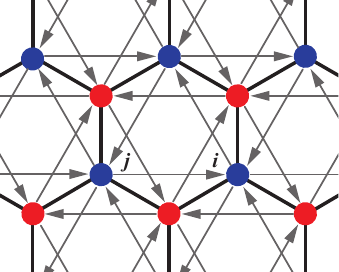}
	\caption{ Schematic of the honeycomb lattice structure. $\nu_{ij}=+1$ when hopping from $j$ to $i$ in the direction of the arrow, otherwise, it has $\nu_{ij}=-1$.}   \label{figure1}
\end{figure}

\begin{figure*}[!tb]	
	\centering
	\includegraphics[width=18cm]{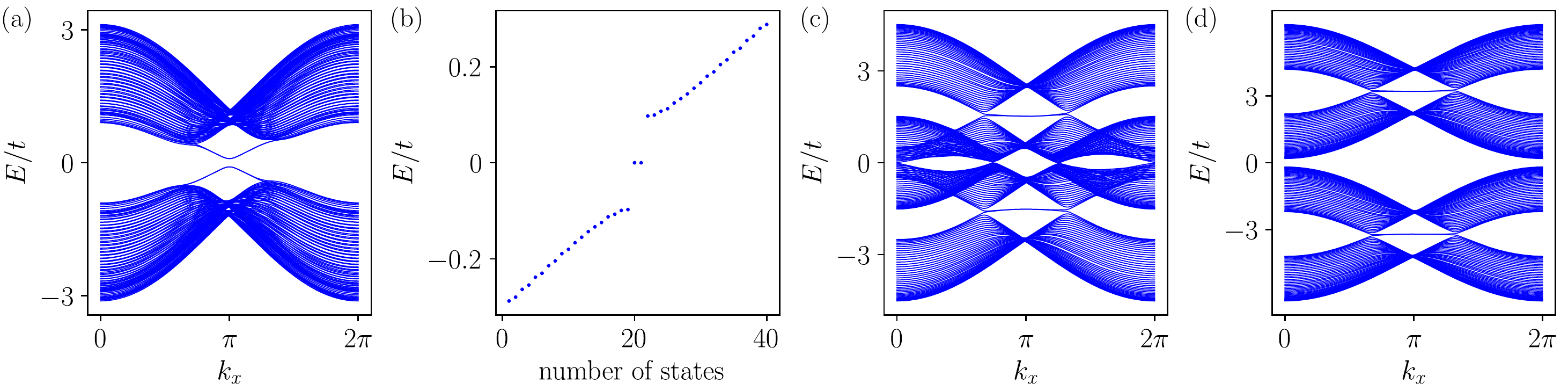}
	\caption{Band structures under open boundary conditions along the $y$ direction for (a) $\lambda/t=0.1$, (c) $\lambda/t=1.5$ and (d) $\lambda/t=3.2$. (b) Eigenenergies under open boundary conditions along the $x$ and $y$ directions for  $\lambda/t=0.1$. } \label{figure2}
\end{figure*}

\section{\label{sec:the model}The Model With Charge-Phonon Coupling}

In order to uncover the effects of electron-phonon interactions on  the higher-order topological properties, we consider the modified  Kane-Mele model\cite{PhysRevLett.124.166804} for spinful electrons in the presence of intrinsic spin-orbit coupling and an in-plane Zeeman field in a honeycomb lattice.  Its tight-binding Hamiltonian is written as\cite{PhysRevLett.124.166804} 
\begin{align}\label{equation1}
	\mathcal{H}_0=-t\sum_{\langle i\,j \rangle}c_{i}^{\dagger}c_{j}+i t_{\mathrm{so}}\sum_{\langle\!\langle i\, j\rangle\!\rangle}\nu_{ij}c_{i}^{\dagger}s_{z}c_{j}+\lambda\sum_{i}c_{i}^{\dagger} s_y c_{i}, 
\end{align}
where $c_{i}^{\dagger}=(c_{i\uparrow}^{\dagger},c_{i\downarrow}^{\dagger})$, with $c_{i\sigma}^{\dagger}$ $(\sigma=\uparrow,\downarrow)$ being the creation operator of an electron with spin-$\sigma$ at the $i$th site, $s_i$ ($i=x,y,z$) is the spin-$1/2$ Pauli matrix, $t$ is the nearest-neighbor hopping  amplitude, $t_{\mathrm{so}}$ denotes the spin-orbit interaction associated with the next-nearest-neighbor hopping with $\nu_{ij}=\pm$, depending on the hopping direction of the electrons [see Fig.~\ref{figure1}(a)], and $\lambda$ represents the in-plane Zeeman field strength along the $y$ direction.  In this paper, we assume $t_{\mathrm{so}}=0.1t$.

In momentum space by Fourier transforming Eq.~(\ref{equation1}), we obtain $H_0 = \sum_\mathbf{k} \Psi^\dagger \mathcal{H}_0(\mathbf{k}) \Psi $, with
\begin{align}\label{equation2}
	\mathcal{H}_0(\mathbf{k})= &\left[t+2 t \cos (\frac{3  k_{y} }{2}) \cos (\frac{\sqrt{3}  k_{x} }{2})\right] \sigma_{x} +\lambda \sigma_{0} s_{y}\nonumber \\
	&+ 2 t \sin (\frac{3  k_{y} }{2}) \cos (\frac{\sqrt{3}  k_{x} }{2}) \sigma_{y}  \nonumber \\
	&+4 t_{\mathrm{so}} \cos (\frac{3  k_{y} }{2}) \sin (\frac{\sqrt{3}  k_{x} }{2})  \sigma_{z} s_{z} 
	\nonumber \\
	&-2 t_{\mathrm{so}} \sin (\sqrt{3}  k_{x}) \sigma_{z} s_{z},
\end{align}
where $\Psi = (c_{A, \mathbf{k},\uparrow}, ~c_{A, \mathbf{k},\downarrow},~c_{B, \mathbf{k},\uparrow}, ~c_{B, \mathbf{k},\downarrow})^T$, and $\sigma$ and $s$ are Pauli matrices acting on the sublattice and spin degrees of freedom, respectively.

The system $\mathcal{H}_0(\mathbf{k})$ in Eq.~(\ref{equation2}) exhibits distinct topological phases, depending on the Zeeman field $\lambda$. In the absence of an in-plane magnetic field (i.e., $\lambda=0$),  $\mathcal{H}_0(\mathbf{k})$ reserves a time-reversal symmetry, and it is a quantum spin Hall insulator characterized by a $\mathbb{Z}_2$ topological number \cite{PhysRevB.98.165101}. This first-order topological phase supports a pair of helical gapless edge modes counterpropagating along the zigzag edge of the lattice. Once we apply the  in-plane magnetic field (i.e., $\lambda \neq 0$), the time-reversal symmetry is broken. In this case, a topological phase transition occurs, and the system enters a second-order topological phase regime. In the open boundary condition, the diamond-shaped nanoflake supports two localized corners modes. These corner modes are protected by the $y$-mirror symmetry $\mathcal{M}_y = i \sigma_{x} s_y$ and chiral symmetry. The existence of second-order corner modes is determined by two mirror-graded winding numbers \cite{PhysRevLett.124.166804} of the Hamiltonian $\mathcal{H}_0(k_x, ~k_y=0)$, defined in the eigenstate subspaces of the $y$-mirror symmetry operator along the high-symmetry line with $ k_y=0 $. 

To be concrete, the system $\mathcal{H}_0(\mathbf{k})$ enters the second-order topological phase regimes for $0<\lambda/t<1$. Figures \ref{figure2}(a) and 2(b) show the eigenenergies under  open boundary conditions along the $y$ direction, and both $x$ and $y$ directions, respectively. Second-order mid gap bound states exist for $\lambda/t=0.1$. For $\lambda/t>1$, the system $\mathcal{H}_0(\mathbf{k})$ shows a topological semimetal phase protected by $y$-mirror symmetry. When considering only the half filling,  it is a topologically trivial semimetal phase for $1<\lambda/t<3$, and a trivial insulator for $\lambda/t>3$.
 
In order to study the effects of electron-phonon coupling on higher-order topological phases, we couple spinful fermions to  the lattice degrees of freedom. The hybrid system is described by the Holstein electron-phonon coupling Hamiltonian\cite{PhysRevLett.123.046401}    
\begin{align}\label{equation3}
H = H_0 + H_\textrm{int},
\end{align}
with
\begin{align}\label{equation4}
H_\textrm{int} =  \omega_{0} \sum_{i} d_{i}^{\dagger} d_{i} +  g \omega_{0} \sum_{i,\sigma}\left(c_{i,\sigma}^{\dagger} c_{i,\sigma}-\frac{1}{2}\right)\left(d_{i}^{\dagger}+d_{i}\right),
\end{align}
where  $d_{i}^{\dagger}$ creates a phonon at site $i$, $\omega_{0}$ is the frequency of the optical phonon mode, and $g$ represents the Holstein electron-phonon coupling strength. In this work, we consider the half filling of spinful electrons, and introduce the dimensionless parameter $\alpha = g^{2} \omega_{0} / 4 t$. By changing the value of $\alpha$, we found that a topological quantum transition from the semimetal phase to the second-order topological insulator phase can occur without changing the parameters of the electronic system.

\section{\label{sec:Charge-phonon coupling} The Theoretical Approaches}

\subsection{High optical-mode-frequency limit and Lang-Firsov transformation}

We first consider a limited case with the phonon-mode frequency $\omega_{0}$ much larger than the electronic parameters, i.e., $\omega_{0} \gg t, ~t_{\textrm{so}}, ~\lambda$, and $g$. Now, we can analytically demonstrate the hidden physics of topological phase transitions caused by electron-phonon coupling. In the high optical-mode-frequency limit, by applying the Lang-Firsov transformation\cite{PhysRevLett.123.046401},  
\begin{align}\label{equation5}
H_{\textrm{LF}}=e^{S} H e^{-S},  
\end{align}
with
\begin{align}\label{equation6} 
S=g \sum_{i,\sigma}\left(c_{i,\sigma}^{\dagger} c_{i,\sigma}-\frac{1}{2}\right)\left(d_{i}^{\dagger}-d_{i}\right),
\end{align}
we obtain, 
\begin{align}\label{equation7} 
	H_{\textrm{LF}} = \tilde{H}_0 + \tilde{H}_\textrm{int},
\end{align}
with
\begin{align}\label{equation8} 
      \tilde{H}_0 =& -t\sum_{\langle i\,j \rangle}c_{i}^{\dagger}c_{j}X^{\dagger}_{i}X_{j}+i t_{\mathrm{so}}\sum_{\langle\!\langle i\, j\rangle\!\rangle}\nu_{ij}c_{i}^{\dagger}s_{z}c_{j}X^{\dagger}_{i}X_{j}\nonumber\\
      &+\lambda\sum_{i}c_{i}^{\dagger} s_y c_{i},
\end{align}
and 
\begin{align}\label{equation9} 
	 \tilde{H}_\textrm{int} =  \omega_{0} \sum_{i} d_{i}^{\dagger} d_{i}-2\omega_{0}g^{2}N_{c},
\end{align}
where $X_{i}=e^{g\left(d_{i}-d_{i}^{\dagger}\right)}$, and $N_{c}$ is the number of unit cells. 

By considering the thermal average of the phonon modes, we obtain $\left\langle X_{i}^{\dagger} X_{j}\right\rangle=e^{-g^{2}\left(2 N_{0}+1\right)}$, where $N_{0}=(e^{\beta \omega_{0}}-1)^{-1}$, with $\beta=1/k_{\textrm{B}}T$, and $k_{\textrm{B}}$ being the Boltzmann constant.  Here we just consider the thermal average of the phonon mode, and neglect its quantum fluctuation  $X_{i}^{\dagger} X_{j}-\left\langle X_{i}^{\dagger} X_{j}\right\rangle$ for the limited condition  $\omega_{0} \gg t, ~t_{\textrm{so}}, ~\lambda$ and $g$. In this antiadiabatic regime, the electronic hopping hardly affects the distribution of phonons, because there exists a larger energy gap between different phonon excitations than the hopping strength of electrons, such that the fast lattice fluctuations make phonon modes immediately follow the charge carriers without modifying their distribution. Therefore, by rewriting Eq.~(\ref{equation8}), the parameters $t$ and $t_\textrm{so}$ are renormalized as
\begin{align}\label{equation10}
\tilde{t}&=te^{-g^{2}\left(2 N_{0}+1\right)}\nonumber\\
\tilde{t}_{\mathrm{so}}&=t_{\mathrm{so}}e^{-g^{2}\left(2 N_{0}+1\right)}.
\end{align}
Note that $\lambda$ in Eq.~(\ref{equation8}) is not renormalized. In this case, when we tune the electron-phonon coupling strength $g$, a topological phase transition can occur.

\subsection{Cluster perturbation theory}

To numerically solve the electron-phonon coupling in Eq.~(\ref{equation3}), we calculate the electronic Green function by employing the cluster perturbation theory  \cite{PhysRevB.66.075129,PhysRevLett.123.046401}.  The basic idea of the cluster perturbation theory is to divided the infinite lattice into identical clusters, and each cluster contains a finite number of lattice sites. In this case, all the clusters form a superlattice structure with each cluster being an unit cell.  In this work, we choose a two-site cluster, containing two nearest-neighbor sublattices, as the unit cell of the superlattice. 

We rewrite Eq.~(\ref{equation3}) as the sum of the cluster Hamiltonian and inter-cluster coupling  $H=H_{0}+V_\textrm{int}$, with 
\begin{align}\label{equation11}
H_{0}=\sum_{\mathbf{r}_c} h_{0}(\mathbf{r}_c),   
\end{align}
and
\begin{align}\label{equation11}
  V_\textrm{int}  = &  -t\sum_{\langle i\,j \rangle} \sum_{\mathbf{r}_c\neq \mathbf{r}_c^{\prime}} c_{i}^{\dagger}(\mathbf{r}_c)c_{j}(\mathbf{r}_c^{\prime}) \nonumber \\
  & +it_{\mathrm{so}}\sum_{\langle\!\langle i\, j\rangle\!\rangle}\sum_{\mathbf{r}_c\neq \mathbf{r}_c^{\prime}}\nu_{ij}c_{i}^{\dagger}(\mathbf{r}_c)s_{z}c_{j}(\mathbf{r}_c^{\prime}),
\end{align}
where $h_{0}(\mathbf{r}_c)$  reads
\begin{figure*}[!tb]	
	\centering
	\includegraphics[width=18cm]{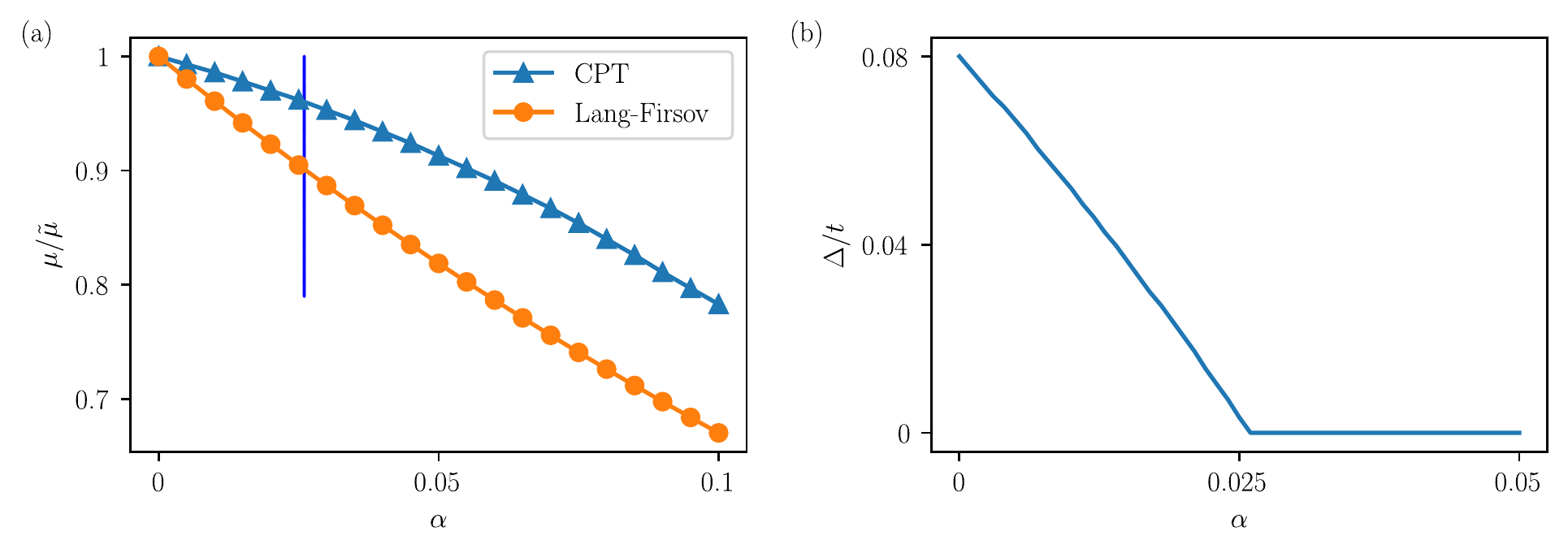}
	\caption{ (a) The renormalized parameters $\mu/\tilde{\mu}$ vs $\alpha$ for the cluster perturbation theory (CPT) and Lang-Firsov approach, respectively. Here, $\mu=\lambda/t$ and $\tilde{\mu}=\tilde{\lambda}/\tilde{t}$. The electron-phonon coupling modifies the system's parameters. The blue vertical line marks the band-gap closing point for $\mu/\tilde{\mu}=0.96$ and $\tilde{\mu}=1$  with $\alpha=g^{2}\omega_{0}/4t=0.026$. (b) The renormalized band gap $\Delta$ versus $\alpha$. For $\alpha = 0.026$, the band gap becomes closed, indicating that a topological phase  transition occurs. }  \label{figure3}
\end{figure*}
\begin{widetext}
\begin{align}\label{equation12}
h_{0}(\mathbf{r}_c) = &-t\sum_{i \neq j}^{2}c_{i}^{\dagger}(\mathbf{r}_c)c_{j}(\mathbf{r}_c)+\lambda\sum_{i}^{2}c_{i}^{\dagger}(\mathbf{r}_c)\mathbf{B}\cdot \mathbf{s}c_{i}(\mathbf{\mathbf{r}_c} )+\omega_{0} \sum_{i} d_{i}^{\dagger}(\mathbf{r}_c) d_{i}(\mathbf{r}_c) \nonumber \\
&+g \omega_{0} \sum_{i}^{2}\left(c_{i}^{\dagger}(\mathbf{r}_c) c_{i}(\mathbf{r}_c)-\frac{1}{2}\right)\left(d_{i}^{\dagger}(\mathbf{r}_c)+d_{i}(\mathbf{r}_c)\right) ,
\end{align}
\end{widetext}
where $h_{0}(\mathbf{r}_c)$ is each cluster Hamiltonian, $N_{c}$ is the number of the clusters, $\mathbf{r}_c$ denotes the  site of the center of each cluster,  the subscripts $i$ and $j$ denote the sites of the element in each cluster, and $V_\textrm{int}$  describes the hopping between clusters. 

When employing the cluster perturbation theory, $V_\textrm{int}$ is considered as a perturbation to the cluster Hamiltonian $H_{0}$. Therefore,   the system Green's function   $G(z)$ can be written in terms of the cluster Green's function $\mathcal{G}(z)$ and the perturbation $V$ as
\begin{align}\label{equation13}
G^{-1}(z)=\mathcal{G}^{-1}(z)-V_\textrm{int}, 
\end{align}
where $z=\omega+i \eta$. 

The original lattice has translation symmetry. Therefore, we can Fourier transform, over the whole superlattice, $G^{-1}(z)$ and $V$ in Eq.~(\ref{equation13}), and achieve 
\begin{align}\label{equation14}
G^{-1}(\tilde{\mathbf{k}},z)=\mathcal{G}^{-1}(z)-V_\textrm{int}(\tilde{\mathbf{k}}), 
\end{align}
where $\tilde{\mathbf{k}}$ is the wave vector corresponding to the first Brillouin zone of the superlattice, and $\tilde{\mathbf{k}} = \mathbf{k}$ for our case. $\mathcal{G}(z)$ is the retarded cluster Green function at zero temperature, and it can be written as
\begin{align}\label{equation15}
\mathcal{G}(z)_{i\sigma ,j\sigma^{\prime}}&=\int dt\ e^{(i\omega-\eta)t}(-i\theta(t)\langle\Omega|[c_{i\sigma}(t),c_{j\sigma^{\prime}}^{\dagger}(0)]_{+}|\Omega\rangle)\nonumber\\
&=\mathcal{G}^{+}_{i\sigma ,j\sigma^{\prime}}(z)+\mathcal{G}^{-}_{i\sigma ,j\sigma^{\prime}}(z), 
\end{align}
where 
\begin{align}\label{equation1511}
\mathcal{G}^{+}_{i\sigma ,j\sigma^{\prime}}(z)=\langle\Omega|c_{i\sigma}(z-h_{0}+E_{0})^{-1}c^{\dagger}_{j\sigma^{\prime}}|\Omega\rangle,  
\end{align}
\begin{align}\label{equation1522}
	\mathcal{G}^{-}_{i\sigma ,j\sigma^{\prime}}(z)=\langle\Omega|c^{\dagger}_{i\sigma}(z+h_{0}-E_{0})^{-1}c_{j\sigma^{\prime}}|\Omega\rangle.  
\end{align}
In Eqs.~(\ref{equation1511}) and (\ref{equation1522}), $|\Omega \rangle$ is the ground state of the cluster Hamiltonian $h_{0}$ in Eq.~(\ref{equation12}), and $E_{0}$ is the corresponding ground-state energy. 

After solving out Eq.~(\ref{equation14}), the one-electron spectral function can be calculated  as
\begin{align}\label{equation16}
A(\tilde{\mathbf{k}},\omega)=-\frac{1}{\pi}\lim _{\eta \rightarrow 0^{+}} \operatorname{Im} {G}(\tilde{\mathbf{k}}, \omega+i \eta),
\end{align}
where $\eta$ can be interpreted as the broadening factor.

\begin{figure*}[!tb]	
	\centering
	\includegraphics[width=18cm]{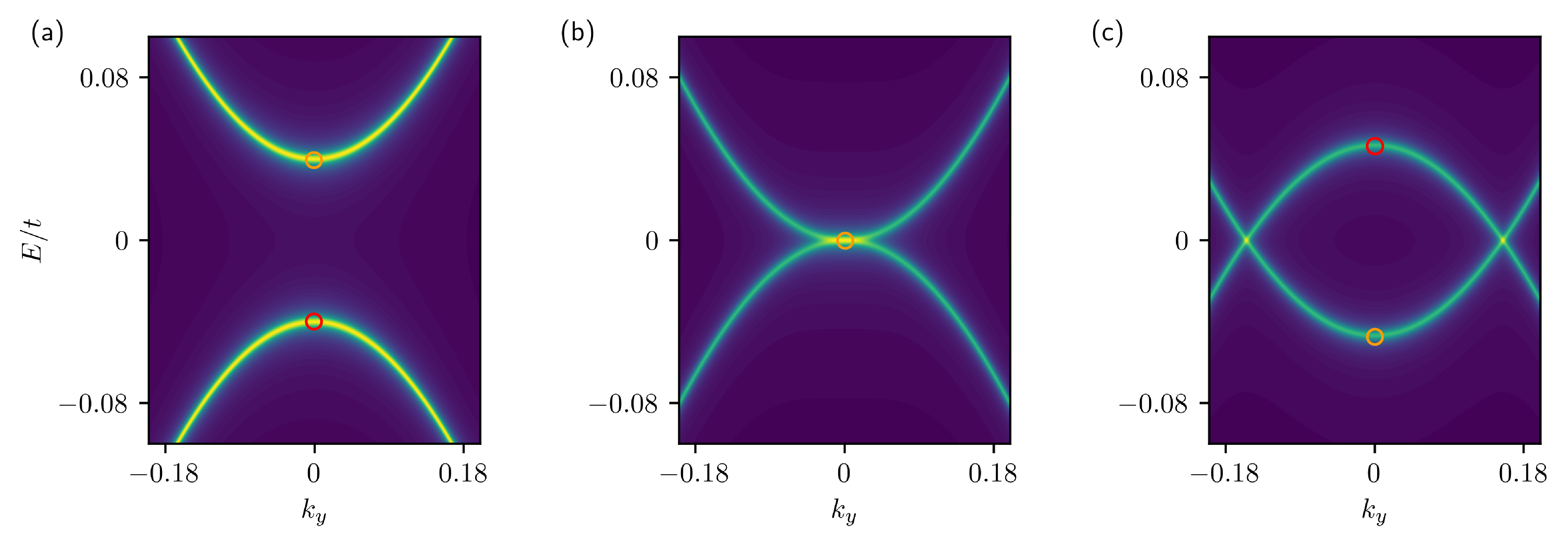}
	\caption{ Energy bands calculated by $\mathcal{H}_\textrm{eff} = -G^{-1}(k_{x}, k_{y},\omega=0)$ with $k_x=0$ for (a) $\alpha=0$, (b) $\alpha=0.026$, and (c) $\alpha=0.05$, respectively. A topological phase transition occurs for $\alpha=0.026$, below which it is an insulator, and becomes a semimetal once $\alpha>0.026$. Here, we only show the two bands closest to the Fermi level at half filling. The bands marked by red and orange circles are used to calculate the fermionic number. Note that the ground states are not degenerate.   }  \label{figure4}
\end{figure*}
\begin{figure*}[!tb]	
	\centering
	\includegraphics[width=18cm]{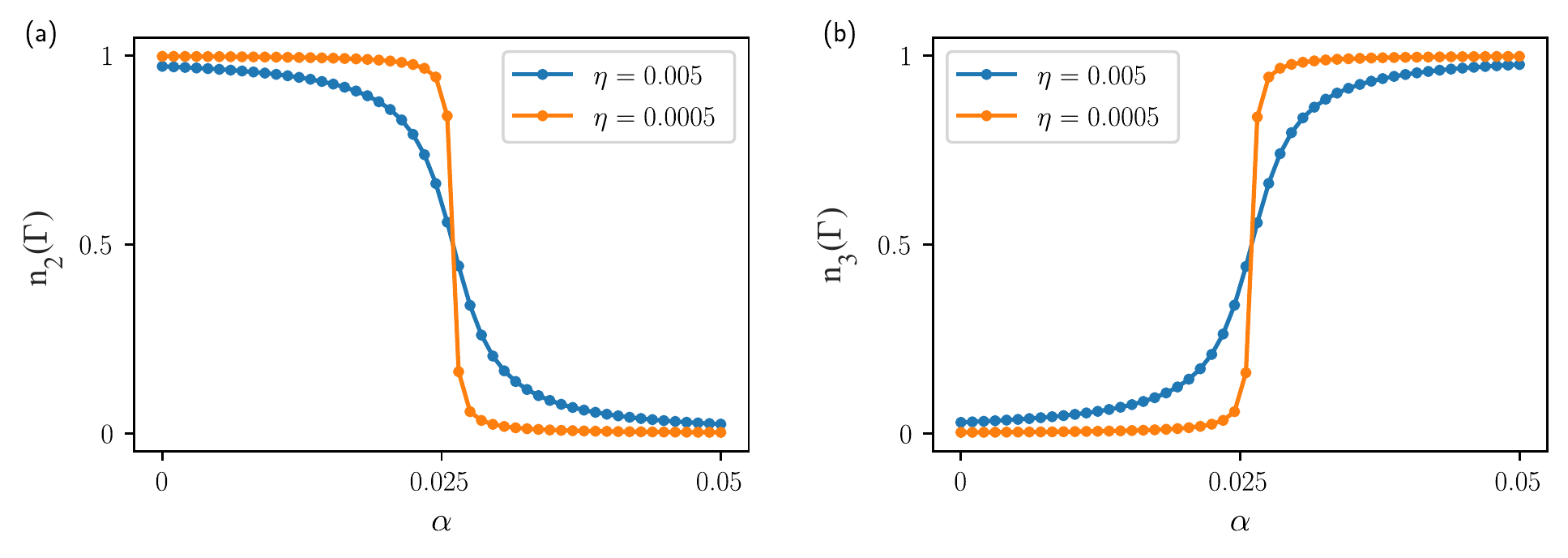}
	\caption{ The average number of fermions $n_{2}(\Gamma)$ and $n_{3}(\Gamma)$ for two bands closest to the Fermi level. These two bands   initially lie below and above the Fermi energy before the electron-phonon coupling is turned on, and then they pass through the Fermi energy as $\alpha$ increases for different broadening factors $\eta$. The subscript here represents the band with the second and third lowest energy when $\alpha$ equals zero.}  \label{figure5}
\end{figure*}
\begin{figure}[!tb]
	\centering
	\includegraphics[width=8.4cm]{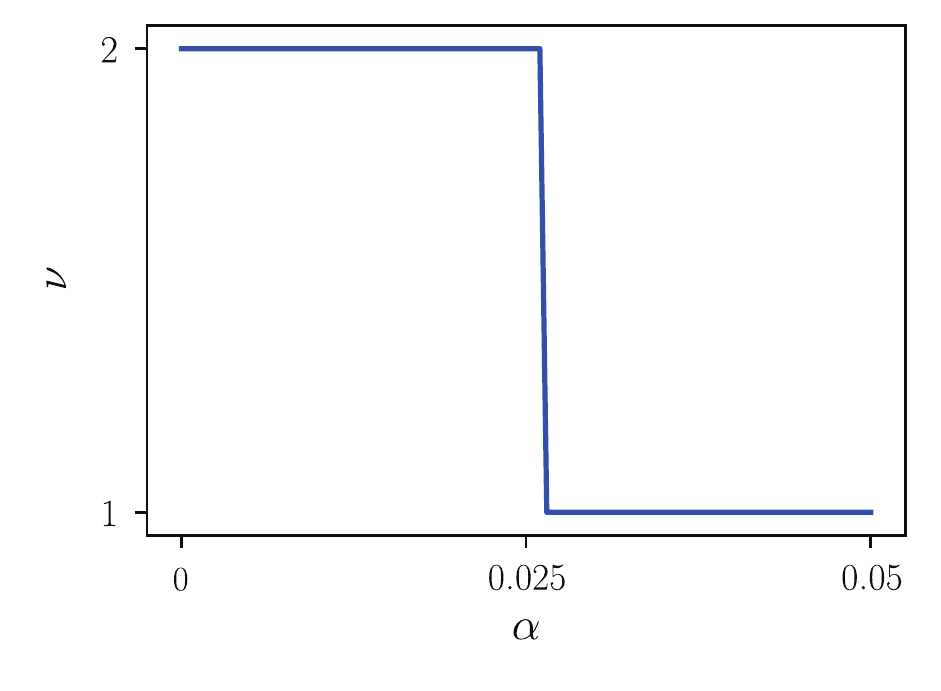}
	\caption{ Winding number $\nu=\nu^{-}+\nu^{+}$ vs $\alpha$. As the $\alpha$ increases, $\nu$ changes from 2 ($\nu^{\pm}=1$) to 1 ($\nu^{+}=0, \nu^{-}=1$), indicating that a second-order topological phase transition occurs.      } 	\label{figure6}
\end{figure}

\begin{figure}[!tb]
	\centering
	\includegraphics[width=8.4cm]{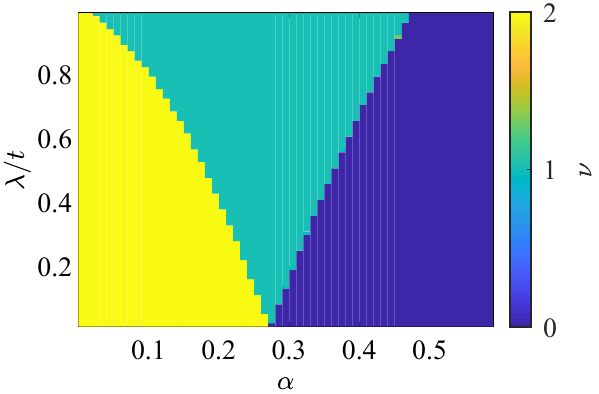}
	\caption{ {\color{red}Phase diagram of the charge-phonon coupling system considered. There exist three phases: a second-order topologically trivial insulator with  $\nu = 0$, a semimetal with $\nu = 1$, and a second-order topologically nontrivial insulator with $\nu = 2$.}     } 	\label{figure7}
\end{figure}

\section{\label{sec:Results} Numerical Results}

To study the effects of an electron-phonon interaction on the higher-order topological features, we set the following parameters, $\omega_{0}/t=1$, $t_{\mathrm{so}}/t=0.1$ and $\lambda/t=0.96$, where the system, in the absence of electron-phonon coupling, is in the second-order topological phase regime near the phase transition point towards a semimetal phase for half-filling. In this work, we consider only the half-filling case.

As stated above, the Lang-Firsov transformation has shown that the electron-phonon coupling can renormalize system parameters. Furthermore, topological phase regimes are determined by $\lambda/t$, as shown in Fig.~\ref{figure2}. To reveal the effect of electron-phonon coupling on the system parameters, we plot the ratio of $\mu=\lambda/t$ to the renormalized value $\tilde{\mu}=\tilde{\lambda}/\tilde{t}$ as an introduced dimensionless parameter $\alpha=g^{2}\omega_{0}/4t$ varies, as shown in Fig.~\ref{figure3}(a). These results calculated by the cluster perturbation theory are compared with those by the Lang-Firsov approach. As $\alpha$ increases,  $\tilde{\mu}$ rises, indicating that the electron-phonon coupling can modify the system's parameters and exhibits the potential for causing the topological phase transitions. In addition, by comparing the results calculated from the Lang-Firsov approach and the ones from the cluster perturbation theory in Fig.~\ref{figure3}(a), as the dimensional parameter $\alpha$ increases, the quantum fluctuation of phonon modes starts playing an important role in determining the electronic properties in the large electron-phonon coupling regime.

The topological phase transitions are accompanied by the band-gap closing. In Fig.~\ref{figure3}(b), we plot the band gaps at half filling versus $\alpha$. As the $\alpha$ increases, the band gap reduces, and become closed for $\alpha = 0.026$, indicating that a topological phase  transition occurs. Furthermore, as shown in Fig.~\ref{figure4}, we calculate the band structures of the effective Hamiltonian $\mathcal{H}_\textrm{eff}$, constructed via the full Green's function at zero frequency  as  $\mathcal{H}_\textrm{eff} = -G^{-1}(k_{x}, k_{y},\omega=0)$, with $k_x=0$ for different $\alpha$, below which it is an insulator, and becomes a semimetal once $\alpha>0.026$. Here we only show the two bands closest to the Fermi level at half filling.  The insulator and semimetal phases are clearly shown, and the phase transition from the insulator to semimetal takes place at $\alpha = 0.026$. Note that the ground states are not degenerate.

To characterize this phase transition, we calculate the average fermionic number $n_{i}$ of the $i$th band at $\Gamma $ point  for half filling as
\begin{align}\label{number}
	n_{i}(k_{x},k_{y})=\int\tilde{A}_{i,i}(k_{x},k_{y},\omega)n_{F}(\omega)d\omega,  
\end{align}
with 
\begin{align}\label{number2}
	\tilde{A}(\mathbf{k},\omega)=\hat{U}^{\dagger}(\mathbf{k})A(\mathbf{k},\omega)\hat{U}(\mathbf{k}),    
\end{align}
where $\hat{U}(\mathbf{k})$ is a unitary matrix used to diagonalize Hamiltonian $\mathcal{H}_{0}(\mathbf{k})$ in Eq.~(\ref{equation2}), and $n_{F}$ is the Fermi-Dirac distribution. In Fig.~\ref{figure5}, we plot $n_{2}(\Gamma)$ and $n_{3}(\Gamma)$ as a function of $\alpha$ respectively at zero temperature. The bands, in Fig.~\ref{figure4}, marked by a red and orange circles are used to calculate the fermionic number.  As $\alpha$ increases, $n_{2}(\Gamma)$ changes from one to zero, while $n_{3}(\Gamma)$ varies from zero to one.  $n_{2}(\Gamma)$ and $n_{3}(\Gamma)$ become discontinuous when we decrease the broadening factor $\eta$ in Eq.~(\ref{equation16}) at the phase transition. This discontinuity indicates the topological phase transitions, which can be directly probed. 

In order to further determine the second-order topological phase transition, we calculate the topological invariant. The bare modified Kane-Mele model, in the presence of electron-phonon coupling, preserves the $y$-mirror symmetry. As in the single-particle case \cite{PhysRevLett.124.166804}, we calculate the winding number of the effective Hamiltonian $\mathcal{H}_\textrm{eff}(k_{x}, k_{y}) = -G^{-1}(k_{x}, k_{y},\omega=0)$\cite{PhysRevB.98.165101}  with $k_y=0$. In the subspace defined $y$-mirror symmetry,  we can rewrite the Hamiltonian $\mathcal{H}_\textrm{eff}$ in block diagonal form along the high symmetry line
\begin{align}\label{equation20}
	\mathcal{H}_\textrm{eff}(k_{x},k_{y}=0)=\begin{pmatrix}
		\mathcal{H}_{+}(k_{x})&0\\
		0&\mathcal{H}_{-}(k_{x})
	\end{pmatrix},
\end{align}
and then we can define the two mirror-graded winding number
\begin{align}\label{equation21}
	\nu^{\pm}=\int_{-\pi}^{\pi}\frac{dk_{x}}{4\pi i}~\text{Tr}~[\tau_{z}\mathcal{H}_{\pm}^{-1}(k_{x})\partial_{k_{x}}\mathcal{H}_{\pm}(k_{x})],
\end{align}
which can indicate a different phase. Note that quantized winding number is ensured by the chiral symmetry of the system.  When the system is in a second-order TI phase, $	\nu^{\pm}=1$, otherwise it is a second-order topological trivial phase. In our work, we directly calculate $\nu=\nu^{-}+\nu^{+}$, which can be written as
\begin{eqnarray}\label{equation23}
\nu=\int \frac{dk_{x}}{4\pi \mathrm{i}}\text { Tr }[\tau_{0}\otimes\tau_{z}G(k_{x},0)\partial_{k_{x}}G^{-1}(k_{x},0)]. 
\end{eqnarray}
If $\nu=2$, the system is in the second-order topological phase, otherwise it is topologically trivial. In Fig.~\ref{figure6}, we plot the winding number as a function of $\alpha$, which is also discontinuous at $\alpha\simeq0.026$. This indicates that a second-order topological phase transition occurs at this point.   {\color{red}In Fig.~\ref{figure7}, we present the phase diagram of the charge-phonon coupling system considered here. There exist three phases: second-order topologically trivial insulator with  $\nu = 0$, a semimetal with $\nu = 1$, and a second-order topologically nontrivial insulator with $\nu = 2$.}  Overall, the electron-phonon coupling studied in this work can induce higher-order topological phase transitions, and the inevitable roles played by such an electron-phonon interaction should be considered in solid-state materials.

\section{\label{sec:Conclusion}Conclusion}
We have considered a modified Kane-Mele model, which hosts a second-order topological phase for the single-particle case, subjected to electron-phonon coupling.  Such a hybrid system is described by the Holstein Hamiltonian, and is solved by using the cluster perturbation approach. We started with the bare modified Kane-Mele model, which is in the second-order topological phase regime in the absence of electron-phonon coupling. When the electron-phonon interaction is turned on, our theoretical calculations  on the spectral function via Green's functions show that the band gap becomes closed as the electron-phonon interaction strength increases, and the fermionic number exhibits a finite discontinuity at the transition point at the critical electron-phonon coupling strength.  These indicates a topological phase transition.  By calculating the topological invariant, the second-order topological phase transition, driven by the electron-phonon coupling, is verified. 

The modified Kane-Mele model can be realized  via applying  ferromagnetism with in-plane anisotropy \cite{PhysRevLett.124.166804} to topological insulators of silicene \cite{PhysRevLett.109.055502} and jacutingaite (Pt$_2$HgSe$_3$) \cite{PhysRevLett.120.117701},	while the electron-phonon coupling may be tuned by the  strain or electric field \cite{PhysRevB.91.245403}. In addition, the electron-phonon coupling studied here may be simulated using Rydberg states of cold atoms and trapped ions\cite{Hague2012,Mendonca, PhysRevLett.128.120404}.

\begin{acknowledgments}
	T.L. acknowledges the support from National Natural Science Foundation of China (Grant No.~12274142) and the Startup Grant of South China University of Technology (Grant No.~20210012). Q.A. is thankful for the support from Beijing Natural Science Foundation (Grant No.~1202017) and Beijing Normal University (Grant No.~2022129). M.Z. acknowledges the support from the National Natural Science Foundation of China (Grant No.~11475021) and the National Key Basic Research Program of China (Grant No.~2013CB922000). H.B.W. is thankful for the support from National Natural Science Foundation of China (Grant No.~61675028) and National Natural Science Foundation of China (Grant No.~12274037)
\end{acknowledgments}


\begin{thebibliography}{71}%
	\makeatletter
	\providecommand \@ifxundefined [1]{%
		\@ifx{#1\undefined}
	}%
	\providecommand \@ifnum [1]{%
		\ifnum #1\expandafter \@firstoftwo
		\else \expandafter \@secondoftwo
		\fi
	}%
	\providecommand \@ifx [1]{%
		\ifx #1\expandafter \@firstoftwo
		\else \expandafter \@secondoftwo
		\fi
	}%
	\providecommand \natexlab [1]{#1}%
	\providecommand \enquote  [1]{``#1''}%
	\providecommand \bibnamefont  [1]{#1}%
	\providecommand \bibfnamefont [1]{#1}%
	\providecommand \citenamefont [1]{#1}%
	\providecommand \href@noop [0]{\@secondoftwo}%
	\providecommand \href [0]{\begingroup \@sanitize@url \@href}%
	\providecommand \@href[1]{\@@startlink{#1}\@@href}%
	\providecommand \@@href[1]{\endgroup#1\@@endlink}%
	\providecommand \@sanitize@url [0]{\catcode `\\12\catcode `\$12\catcode
		`\&12\catcode `\#12\catcode `\^12\catcode `\_12\catcode `\%12\relax}%
	\providecommand \@@startlink[1]{}%
	\providecommand \@@endlink[0]{}%
	\providecommand \url  [0]{\begingroup\@sanitize@url \@url }%
	\providecommand \@url [1]{\endgroup\@href {#1}{\urlprefix }}%
	\providecommand \urlprefix  [0]{URL }%
	\providecommand \Eprint [0]{\href }%
	\providecommand \doibase [0]{http://dx.doi.org/}%
	\providecommand \selectlanguage [0]{\@gobble}%
	\providecommand \bibinfo  [0]{\@secondoftwo}%
	\providecommand \bibfield  [0]{\@secondoftwo}%
	\providecommand \translation [1]{[#1]}%
	\providecommand \BibitemOpen [0]{}%
	\providecommand \bibitemStop [0]{}%
	\providecommand \bibitemNoStop [0]{.\EOS\space}%
	\providecommand \EOS [0]{\spacefactor3000\relax}%
	\providecommand \BibitemShut  [1]{\csname bibitem#1\endcsname}%
	\let\auto@bib@innerbib\@empty
	\bibitem [{\citenamefont {Zhang}\ \emph {et~al.}(2013)\citenamefont {Zhang},
		\citenamefont {Kane},\ and\ \citenamefont {Mele}}]{PhysRevLett.110.046404}%
	\BibitemOpen
	\bibfield  {author} {\bibinfo {author} {\bibfnamefont {F.}~\bibnamefont
			{Zhang}}, \bibinfo {author} {\bibfnamefont {C.~L.}\ \bibnamefont {Kane}}, \
		and\ \bibinfo {author} {\bibfnamefont {E.~J.}\ \bibnamefont {Mele}},\
	}\bibfield  {title} {\enquote {\bibinfo {title} {Surface state magnetization
				and chiral edge states on topological insulators},}\ }\href
	{https://link.aps.org/doi/10.1103/PhysRevLett.110.046404} {\bibfield
		{journal} {\bibinfo  {journal} {Phys. Rev. Lett.}\ }\textbf {\bibinfo
			{volume} {110}},\ \bibinfo {pages} {046404} (\bibinfo {year}
		{2013})}\BibitemShut {NoStop}%
	\bibitem [{\citenamefont {Benalcazar}\ \emph
		{et~al.}(2017{\natexlab{a}})\citenamefont {Benalcazar}, \citenamefont
		{Bernevig},\ and\ \citenamefont {Hughes}}]{Benalcazar61}%
	\BibitemOpen
	\bibfield  {author} {\bibinfo {author} {\bibfnamefont {W.~A.}\ \bibnamefont
			{Benalcazar}}, \bibinfo {author} {\bibfnamefont {B.~A.}\ \bibnamefont
			{Bernevig}}, \ and\ \bibinfo {author} {\bibfnamefont {T.~L.}\ \bibnamefont
			{Hughes}},\ }\bibfield  {title} {\enquote {\bibinfo {title} {Quantized
				electric multipole insulators},}\ }\href
	{http://science.sciencemag.org/content/359/6371/76.abstract} {\bibfield
		{journal} {\bibinfo  {journal} {Science}\ }\textbf {\bibinfo {volume}
			{357}},\ \bibinfo {pages} {61} (\bibinfo {year}
		{2017}{\natexlab{a}})}\BibitemShut {NoStop}%
	\bibitem [{\citenamefont {Benalcazar}\ \emph
		{et~al.}(2017{\natexlab{b}})\citenamefont {Benalcazar}, \citenamefont
		{Bernevig},\ and\ \citenamefont {Hughes}}]{PhysRevB.96.245115}%
	\BibitemOpen
	\bibfield  {author} {\bibinfo {author} {\bibfnamefont {W.~A.}\ \bibnamefont
			{Benalcazar}}, \bibinfo {author} {\bibfnamefont {B.~A.}\ \bibnamefont
			{Bernevig}}, \ and\ \bibinfo {author} {\bibfnamefont {T.~L.}\ \bibnamefont
			{Hughes}},\ }\bibfield  {title} {\enquote {\bibinfo {title} {Electric
				multipole moments, topological multipole moment pumping, and chiral hinge
				states in crystalline insulators},}\ }\href {\doibase
		10.1103/PhysRevB.96.245115} {\bibfield  {journal} {\bibinfo  {journal} {Phys.
				Rev. B}\ }\textbf {\bibinfo {volume} {96}},\ \bibinfo {pages} {245115}
		(\bibinfo {year} {2017}{\natexlab{b}})}\BibitemShut {NoStop}%
	\bibitem [{\citenamefont {Langbehn}\ \emph {et~al.}(2017)\citenamefont
		{Langbehn}, \citenamefont {Peng}, \citenamefont {Trifunovic}, \citenamefont
		{von Oppen},\ and\ \citenamefont {Brouwer}}]{PhysRevLett.119.246401}%
	\BibitemOpen
	\bibfield  {author} {\bibinfo {author} {\bibfnamefont {J.}~\bibnamefont
			{Langbehn}}, \bibinfo {author} {\bibfnamefont {Y.}~\bibnamefont {Peng}},
		\bibinfo {author} {\bibfnamefont {L.}~\bibnamefont {Trifunovic}}, \bibinfo
		{author} {\bibfnamefont {F.}~\bibnamefont {von Oppen}}, \ and\ \bibinfo
		{author} {\bibfnamefont {P.~W.}\ \bibnamefont {Brouwer}},\ }\bibfield
	{title} {\enquote {\bibinfo {title} {Reflection-symmetric second-order
				topological insulators and superconductors},}\ }\href
	{https://link.aps.org/doi/10.1103/PhysRevLett.119.246401} {\bibfield
		{journal} {\bibinfo  {journal} {Phys. Rev. Lett.}\ }\textbf {\bibinfo
			{volume} {119}},\ \bibinfo {pages} {246401} (\bibinfo {year}
		{2017})}\BibitemShut {NoStop}%
	\bibitem [{\citenamefont {Song}\ \emph {et~al.}(2017)\citenamefont {Song},
		\citenamefont {Fang},\ and\ \citenamefont {Fang}}]{PhysRevLett.119.246402}%
	\BibitemOpen
	\bibfield  {author} {\bibinfo {author} {\bibfnamefont {Z.~D.}\ \bibnamefont
			{Song}}, \bibinfo {author} {\bibfnamefont {Z.}~\bibnamefont {Fang}}, \ and\
		\bibinfo {author} {\bibfnamefont {C.}~\bibnamefont {Fang}},\ }\bibfield
	{title} {\enquote {\bibinfo {title} {$(d\ensuremath{-}2)$-dimensional edge
				states of rotation symmetry protected topological states},}\ }\href {\doibase
		10.1103/PhysRevLett.119.246402} {\bibfield  {journal} {\bibinfo  {journal}
			{Phys. Rev. Lett.}\ }\textbf {\bibinfo {volume} {119}},\ \bibinfo {pages}
		{246402} (\bibinfo {year} {2017})}\BibitemShut {NoStop}%
	\bibitem [{\citenamefont {Kunst}\ \emph {et~al.}(2018)\citenamefont {Kunst},
		\citenamefont {van Miert},\ and\ \citenamefont
		{Bergholtz}}]{PhysRevB.97.241405}%
	\BibitemOpen
	\bibfield  {author} {\bibinfo {author} {\bibfnamefont {F.~K.}\ \bibnamefont
			{Kunst}}, \bibinfo {author} {\bibfnamefont {G.}~\bibnamefont {van Miert}}, \
		and\ \bibinfo {author} {\bibfnamefont {E.~J.}\ \bibnamefont {Bergholtz}},\
	}\bibfield  {title} {\enquote {\bibinfo {title} {Lattice models with exactly
				solvable topological hinge and corner states},}\ }\href
	{https://link.aps.org/doi/10.1103/PhysRevB.97.241405} {\bibfield  {journal}
		{\bibinfo  {journal} {Phys. Rev. B}\ }\textbf {\bibinfo {volume} {97}},\
		\bibinfo {pages} {241405} (\bibinfo {year} {2018})}\BibitemShut {NoStop}%
	\bibitem [{\citenamefont {Peterson}\ \emph {et~al.}(2018)\citenamefont
		{Peterson}, \citenamefont {Benalcazar}, \citenamefont {Hughes},\ and\
		\citenamefont {Bahl}}]{peterson2018quantized}%
	\BibitemOpen
	\bibfield  {author} {\bibinfo {author} {\bibfnamefont {C.~W.}\ \bibnamefont
			{Peterson}}, \bibinfo {author} {\bibfnamefont {W.~A.}\ \bibnamefont
			{Benalcazar}}, \bibinfo {author} {\bibfnamefont {T.~L.}\ \bibnamefont
			{Hughes}}, \ and\ \bibinfo {author} {\bibfnamefont {G.}~\bibnamefont
			{Bahl}},\ }\bibfield  {title} {\enquote {\bibinfo {title} {A quantized
				microwave quadrupole insulator with topologically protected corner states},}\
	}\href {https://doi.org/10.1038/nature25777} {\bibfield  {journal} {\bibinfo
			{journal} {Nature}\ }\textbf {\bibinfo {volume} {555}},\ \bibinfo {pages}
		{346--350} (\bibinfo {year} {2018})}\BibitemShut {NoStop}%
	\bibitem [{\citenamefont {Serra-Garcia}\ \emph {et~al.}(2018)\citenamefont
		{Serra-Garcia}, \citenamefont {Peri}, \citenamefont {S{\"u}sstrunk},
		\citenamefont {Bilal}, \citenamefont {Larsen}, \citenamefont {Villanueva},\
		and\ \citenamefont {Huber}}]{serra2018observation}%
	\BibitemOpen
	\bibfield  {author} {\bibinfo {author} {\bibfnamefont {M.}~\bibnamefont
			{Serra-Garcia}}, \bibinfo {author} {\bibfnamefont {V.}~\bibnamefont {Peri}},
		\bibinfo {author} {\bibfnamefont {R.}~\bibnamefont {S{\"u}sstrunk}}, \bibinfo
		{author} {\bibfnamefont {O.~R.}\ \bibnamefont {Bilal}}, \bibinfo {author}
		{\bibfnamefont {T.}~\bibnamefont {Larsen}}, \bibinfo {author} {\bibfnamefont
			{L.~G.}\ \bibnamefont {Villanueva}}, \ and\ \bibinfo {author} {\bibfnamefont
			{S.~D.}\ \bibnamefont {Huber}},\ }\bibfield  {title} {\enquote {\bibinfo
			{title} {Observation of a phononic quadrupole topological insulator},}\
	}\href {\doibase https://doi.org/10.1038/nature25156} {\bibfield  {journal}
		{\bibinfo  {journal} {Nature}\ }\textbf {\bibinfo {volume} {555}},\ \bibinfo
		{pages} {342--345} (\bibinfo {year} {2018})}\BibitemShut {NoStop}%
	\bibitem [{\citenamefont {Geier}\ \emph {et~al.}(2018)\citenamefont {Geier},
		\citenamefont {Trifunovic}, \citenamefont {Hoskam},\ and\ \citenamefont
		{Brouwer}}]{arXiv:1801.10053}%
	\BibitemOpen
	\bibfield  {author} {\bibinfo {author} {\bibfnamefont {M.}~\bibnamefont
			{Geier}}, \bibinfo {author} {\bibfnamefont {L.}~\bibnamefont {Trifunovic}},
		\bibinfo {author} {\bibfnamefont {M.}~\bibnamefont {Hoskam}}, \ and\ \bibinfo
		{author} {\bibfnamefont {P.~W.}\ \bibnamefont {Brouwer}},\ }\bibfield
	{title} {\enquote {\bibinfo {title} {Second-order topological insulators and
				superconductors with an order-two crystalline symmetry},}\ }\href
	{https://link.aps.org/doi/10.1103/PhysRevB.97.205135} {\bibfield  {journal}
		{\bibinfo  {journal} {Phys. Rev. B}\ }\textbf {\bibinfo {volume} {97}},\
		\bibinfo {pages} {205135} (\bibinfo {year} {2018})}\BibitemShut {NoStop}%
	\bibitem [{\citenamefont {Ezawa}(2018)}]{PhysRevLett.120.026801}%
	\BibitemOpen
	\bibfield  {author} {\bibinfo {author} {\bibfnamefont {M.}~\bibnamefont
			{Ezawa}},\ }\bibfield  {title} {\enquote {\bibinfo {title} {Higher-order
				topological insulators and semimetals on the breathing kagome and pyrochlore
				lattices},}\ }\href {https://link.aps.org/doi/10.1103/PhysRevLett.120.026801}
	{\bibfield  {journal} {\bibinfo  {journal} {Phys. Rev. Lett.}\ }\textbf
		{\bibinfo {volume} {120}},\ \bibinfo {pages} {026801} (\bibinfo {year}
		{2018})}\BibitemShut {NoStop}%
	\bibitem [{\citenamefont {Schindler}\ \emph
		{et~al.}(2018{\natexlab{a}})\citenamefont {Schindler}, \citenamefont {Cook},
		\citenamefont {Vergniory}, \citenamefont {Wang}, \citenamefont {Parkin},
		\citenamefont {Bernevig},\ and\ \citenamefont
		{Neupert}}]{schindler2018higher}%
	\BibitemOpen
	\bibfield  {author} {\bibinfo {author} {\bibfnamefont {F.}~\bibnamefont
			{Schindler}}, \bibinfo {author} {\bibfnamefont {A.~M.}\ \bibnamefont {Cook}},
		\bibinfo {author} {\bibfnamefont {M.~G.}\ \bibnamefont {Vergniory}}, \bibinfo
		{author} {\bibfnamefont {Z.~J.}\ \bibnamefont {Wang}}, \bibinfo {author}
		{\bibfnamefont {S.~S.}\ \bibnamefont {Parkin}}, \bibinfo {author}
		{\bibfnamefont {B.~A.}\ \bibnamefont {Bernevig}}, \ and\ \bibinfo {author}
		{\bibfnamefont {T.}~\bibnamefont {Neupert}},\ }\bibfield  {title} {\enquote
		{\bibinfo {title} {Higher-order topological insulators},}\ }\href
	{https://www.science.org/doi/10.1126/sciadv.aat0346} {\bibfield  {journal}
		{\bibinfo  {journal} {Science advances}\ }\textbf {\bibinfo {volume} {4}},\
		\bibinfo {pages} {eaat0346} (\bibinfo {year}
		{2018}{\natexlab{a}})}\BibitemShut {NoStop}%
	\bibitem [{\citenamefont {Zhang}\ \emph {et~al.}(2019)\citenamefont {Zhang},
		\citenamefont {Wang}, \citenamefont {Lin}, \citenamefont {Tian},
		\citenamefont {Xie}, \citenamefont {Lu}, \citenamefont {Chen},\ and\
		\citenamefont {Jiang}}]{Zhang2019}%
	\BibitemOpen
	\bibfield  {author} {\bibinfo {author} {\bibfnamefont {X.}~\bibnamefont
			{Zhang}}, \bibinfo {author} {\bibfnamefont {H.~X.}\ \bibnamefont {Wang}},
		\bibinfo {author} {\bibfnamefont {Z.~K.}\ \bibnamefont {Lin}}, \bibinfo
		{author} {\bibfnamefont {Y.}~\bibnamefont {Tian}}, \bibinfo {author}
		{\bibfnamefont {B.}~\bibnamefont {Xie}}, \bibinfo {author} {\bibfnamefont
			{M.~H.}\ \bibnamefont {Lu}}, \bibinfo {author} {\bibfnamefont {Y.~F.}\
			\bibnamefont {Chen}}, \ and\ \bibinfo {author} {\bibfnamefont {J.~H.}\
			\bibnamefont {Jiang}},\ }\bibfield  {title} {\enquote {\bibinfo {title}
			{Second-order topology and multidimensional topological transitions in sonic
				crystals},}\ }\href {\doibase 10.1038/s41567-019-0472-1} {\bibfield
		{journal} {\bibinfo  {journal} {Nat. Phys.}\ }\textbf {\bibinfo {volume}
			{15}},\ \bibinfo {pages} {582} (\bibinfo {year} {2019})}\BibitemShut
	{NoStop}%
	\bibitem [{\citenamefont {Ni}\ \emph {et~al.}(2019)\citenamefont {Ni},
		\citenamefont {Weiner}, \citenamefont {Al{\`{u}}},\ and\ \citenamefont
		{Khanikaev}}]{Ni2018}%
	\BibitemOpen
	\bibfield  {author} {\bibinfo {author} {\bibfnamefont {X.}~\bibnamefont
			{Ni}}, \bibinfo {author} {\bibfnamefont {M.}~\bibnamefont {Weiner}}, \bibinfo
		{author} {\bibfnamefont {A.}~\bibnamefont {Al{\`{u}}}}, \ and\ \bibinfo
		{author} {\bibfnamefont {A.~B.}\ \bibnamefont {Khanikaev}},\ }\bibfield
	{title} {\enquote {\bibinfo {title} {Observation of higher-order topological
				acoustic states protected by generalized chiral symmetry},}\ }\href {\doibase
		10.1038/s41563-018-0252-9} {\bibfield  {journal} {\bibinfo  {journal} {Nat.
				Mater.}\ }\textbf {\bibinfo {volume} {18}},\ \bibinfo {pages} {113} (\bibinfo
		{year} {2019})}\BibitemShut {NoStop}%
	\bibitem [{\citenamefont {Xue}\ \emph {et~al.}(2019)\citenamefont {Xue},
		\citenamefont {Yang}, \citenamefont {Gao}, \citenamefont {Chong},\ and\
		\citenamefont {Zhang}}]{xue2019acoustic}%
	\BibitemOpen
	\bibfield  {author} {\bibinfo {author} {\bibfnamefont {H.}~\bibnamefont
			{Xue}}, \bibinfo {author} {\bibfnamefont {Y.}~\bibnamefont {Yang}}, \bibinfo
		{author} {\bibfnamefont {F.}~\bibnamefont {Gao}}, \bibinfo {author}
		{\bibfnamefont {Y.~D.}\ \bibnamefont {Chong}}, \ and\ \bibinfo {author}
		{\bibfnamefont {B.}~\bibnamefont {Zhang}},\ }\bibfield  {title} {\enquote
		{\bibinfo {title} {Acoustic higher-order topological insulator on a kagome
				lattice},}\ }\href {\doibase https://doi.org/10.1038/s41563-018-0251-x}
	{\bibfield  {journal} {\bibinfo  {journal} {Nature materials}\ }\textbf
		{\bibinfo {volume} {18}},\ \bibinfo {pages} {108--112} (\bibinfo {year}
		{2019})}\BibitemShut {NoStop}%
	\bibitem [{\citenamefont {Khalaf}(2018)}]{arXiv:1801.10050}%
	\BibitemOpen
	\bibfield  {author} {\bibinfo {author} {\bibfnamefont {E.}~\bibnamefont
			{Khalaf}},\ }\bibfield  {title} {\enquote {\bibinfo {title} {Higher-order
				topological insulators and superconductors protected by inversion
				symmetry},}\ }\href {https://link.aps.org/doi/10.1103/PhysRevB.97.205136}
	{\bibfield  {journal} {\bibinfo  {journal} {Phys. Rev. B}\ }\textbf {\bibinfo
			{volume} {97}},\ \bibinfo {pages} {205136} (\bibinfo {year}
		{2018})}\BibitemShut {NoStop}%
	\bibitem [{\citenamefont {Schindler}\ \emph
		{et~al.}(2018{\natexlab{b}})\citenamefont {Schindler}, \citenamefont {Wang},
		\citenamefont {Vergniory}, \citenamefont {Cook}, \citenamefont {Murani},
		\citenamefont {Sengupta}, \citenamefont {Kasumov}, \citenamefont {Deblock},
		\citenamefont {Drozdov}, \citenamefont {Bouchiat}, \citenamefont {Guéron},
		\citenamefont {Yazdani}, \citenamefont {Bernevig},\ and\ \citenamefont
		{Neupert}}]{arXiv:1802.02585}%
	\BibitemOpen
	\bibfield  {author} {\bibinfo {author} {\bibfnamefont {F.}~\bibnamefont
			{Schindler}}, \bibinfo {author} {\bibfnamefont {Z.}~\bibnamefont {Wang}},
		\bibinfo {author} {\bibfnamefont {M.~G.}\ \bibnamefont {Vergniory}}, \bibinfo
		{author} {\bibfnamefont {A.~M.}\ \bibnamefont {Cook}}, \bibinfo {author}
		{\bibfnamefont {A.}~\bibnamefont {Murani}}, \bibinfo {author} {\bibfnamefont
			{S.}~\bibnamefont {Sengupta}}, \bibinfo {author} {\bibfnamefont {A.~Y.}\
			\bibnamefont {Kasumov}}, \bibinfo {author} {\bibfnamefont {R.}~\bibnamefont
			{Deblock}}, \bibinfo {author} {\bibfnamefont {S.~J.~I.}\ \bibnamefont
			{Drozdov}}, \bibinfo {author} {\bibfnamefont {H.}~\bibnamefont {Bouchiat}},
		\bibinfo {author} {\bibfnamefont {S.}~\bibnamefont {Guéron}}, \bibinfo
		{author} {\bibfnamefont {A.}~\bibnamefont {Yazdani}}, \bibinfo {author}
		{\bibfnamefont {B.~A.}\ \bibnamefont {Bernevig}}, \ and\ \bibinfo {author}
		{\bibfnamefont {T.}~\bibnamefont {Neupert}},\ }\bibfield  {title} {\enquote
		{\bibinfo {title} {Higher-order topology in \uppercase{B}ismuth},}\ }\href
	{https://www.nature.com/articles/s41567-018-0224-7} {\bibfield  {journal}
		{\bibinfo  {journal} {Nat. Phys.}\ }\textbf {\bibinfo {volume} {14}},\
		\bibinfo {pages} {918} (\bibinfo {year} {2018}{\natexlab{b}})}\BibitemShut
	{NoStop}%
	\bibitem [{\citenamefont {Park}\ \emph {et~al.}(2019)\citenamefont {Park},
		\citenamefont {Kim}, \citenamefont {Cho},\ and\ \citenamefont
		{Lee}}]{PhysRevLett.123.216803}%
	\BibitemOpen
	\bibfield  {author} {\bibinfo {author} {\bibfnamefont {M.~J.}\ \bibnamefont
			{Park}}, \bibinfo {author} {\bibfnamefont {Y.}~\bibnamefont {Kim}}, \bibinfo
		{author} {\bibfnamefont {G.~Y.}\ \bibnamefont {Cho}}, \ and\ \bibinfo
		{author} {\bibfnamefont {S.}~\bibnamefont {Lee}},\ }\bibfield  {title}
	{\enquote {\bibinfo {title} {Higher-order topological insulator in twisted
				bilayer graphene},}\ }\href {\doibase 10.1103/PhysRevLett.123.216803}
	{\bibfield  {journal} {\bibinfo  {journal} {Phys. Rev. Lett.}\ }\textbf
		{\bibinfo {volume} {123}},\ \bibinfo {pages} {216803} (\bibinfo {year}
		{2019})}\BibitemShut {NoStop}%
	\bibitem [{\citenamefont {Yang}\ \emph {et~al.}(2020)\citenamefont {Yang},
		\citenamefont {Li}, \citenamefont {Duan},\ and\ \citenamefont
		{Xu}}]{PhysRevResearch.2.033029}%
	\BibitemOpen
	\bibfield  {author} {\bibinfo {author} {\bibfnamefont {Y.~B.}\ \bibnamefont
			{Yang}}, \bibinfo {author} {\bibfnamefont {K.}~\bibnamefont {Li}}, \bibinfo
		{author} {\bibfnamefont {L.-M.}\ \bibnamefont {Duan}}, \ and\ \bibinfo
		{author} {\bibfnamefont {Y.}~\bibnamefont {Xu}},\ }\bibfield  {title}
	{\enquote {\bibinfo {title} {Type-{II} quadrupole topological insulators},}\
	}\href {\doibase 10.1103/PhysRevResearch.2.033029} {\bibfield  {journal}
		{\bibinfo  {journal} {Phys. Rev. Research}\ }\textbf {\bibinfo {volume}
			{2}},\ \bibinfo {pages} {033029} (\bibinfo {year} {2020})}\BibitemShut
	{NoStop}%
	\bibitem [{\citenamefont {Chen}\ \emph {et~al.}(2020)\citenamefont {Chen},
		\citenamefont {Chen}, \citenamefont {Gao}, \citenamefont {Zhou},\ and\
		\citenamefont {Xu}}]{PhysRevLett.124.036803}%
	\BibitemOpen
	\bibfield  {author} {\bibinfo {author} {\bibfnamefont {R.}~\bibnamefont
			{Chen}}, \bibinfo {author} {\bibfnamefont {C.~Z.}\ \bibnamefont {Chen}},
		\bibinfo {author} {\bibfnamefont {J.~H.}\ \bibnamefont {Gao}}, \bibinfo
		{author} {\bibfnamefont {B.}~\bibnamefont {Zhou}}, \ and\ \bibinfo {author}
		{\bibfnamefont {D.~H.}\ \bibnamefont {Xu}},\ }\bibfield  {title} {\enquote
		{\bibinfo {title} {Higher-order topological insulators in quasicrystals},}\
	}\href {\doibase 10.1103/PhysRevLett.124.036803} {\bibfield  {journal}
		{\bibinfo  {journal} {Phys. Rev. Lett.}\ }\textbf {\bibinfo {volume} {124}},\
		\bibinfo {pages} {036803} (\bibinfo {year} {2020})}\BibitemShut {NoStop}%
	\bibitem [{\citenamefont {Zeng}\ \emph {et~al.}(2020)\citenamefont {Zeng},
		\citenamefont {Yang},\ and\ \citenamefont {Xu}}]{PhysRevB.101.241104}%
	\BibitemOpen
	\bibfield  {author} {\bibinfo {author} {\bibfnamefont {Q.~B.}\ \bibnamefont
			{Zeng}}, \bibinfo {author} {\bibfnamefont {Y.~B.}\ \bibnamefont {Yang}}, \
		and\ \bibinfo {author} {\bibfnamefont {Y.}~\bibnamefont {Xu}},\ }\bibfield
	{title} {\enquote {\bibinfo {title} {Higher-order topological insulators and
				semimetals in generalized {A}ubry-{A}ndr\'e-{H}arper models},}\ }\href
	{\doibase 10.1103/PhysRevB.101.241104} {\bibfield  {journal} {\bibinfo
			{journal} {Phys. Rev. B}\ }\textbf {\bibinfo {volume} {101}},\ \bibinfo
		{pages} {241104} (\bibinfo {year} {2020})}\BibitemShut {NoStop}%
	\bibitem [{\citenamefont {Banerjee}\ \emph {et~al.}(2020)\citenamefont
		{Banerjee}, \citenamefont {Mandal},\ and\ \citenamefont
		{Liew}}]{PhysRevLett.124.063901}%
	\BibitemOpen
	\bibfield  {author} {\bibinfo {author} {\bibfnamefont {R.}~\bibnamefont
			{Banerjee}}, \bibinfo {author} {\bibfnamefont {S.}~\bibnamefont {Mandal}}, \
		and\ \bibinfo {author} {\bibfnamefont {T.~C.~H.}\ \bibnamefont {Liew}},\
	}\bibfield  {title} {\enquote {\bibinfo {title} {Coupling between
				exciton-polariton corner modes through edge states},}\ }\href {\doibase
		10.1103/PhysRevLett.124.063901} {\bibfield  {journal} {\bibinfo  {journal}
			{Phys. Rev. Lett.}\ }\textbf {\bibinfo {volume} {124}},\ \bibinfo {pages}
		{063901} (\bibinfo {year} {2020})}\BibitemShut {NoStop}%
	\bibitem [{\citenamefont {Liu}\ \emph {et~al.}(2021{\natexlab{a}})\citenamefont
		{Liu}, \citenamefont {Hu}, \citenamefont {Chen}, \citenamefont {Zhou},\ and\
		\citenamefont {Xu}}]{PhysRevB.103.L201115}%
	\BibitemOpen
	\bibfield  {author} {\bibinfo {author} {\bibfnamefont {Z.~R.}\ \bibnamefont
			{Liu}}, \bibinfo {author} {\bibfnamefont {L.~H.}\ \bibnamefont {Hu}},
		\bibinfo {author} {\bibfnamefont {C.~Z.}\ \bibnamefont {Chen}}, \bibinfo
		{author} {\bibfnamefont {B.}~\bibnamefont {Zhou}}, \ and\ \bibinfo {author}
		{\bibfnamefont {D.~H.}\ \bibnamefont {Xu}},\ }\bibfield  {title} {\enquote
		{\bibinfo {title} {Topological excitonic corner states and nodal phase in
				bilayer quantum spin hall insulators},}\ }\href {\doibase
		10.1103/PhysRevB.103.L201115} {\bibfield  {journal} {\bibinfo  {journal}
			{Phys. Rev. B}\ }\textbf {\bibinfo {volume} {103}},\ \bibinfo {pages}
		{L201115} (\bibinfo {year} {2021}{\natexlab{a}})}\BibitemShut {NoStop}%
	\bibitem [{\citenamefont {Hua}\ \emph {et~al.}(2022)\citenamefont {Hua},
		\citenamefont {Xiao}, \citenamefont {Liu}, \citenamefont {Sun}, \citenamefont
		{Gao}, \citenamefont {Chen}, \citenamefont {Tong}, \citenamefont {Zhou},\
		and\ \citenamefont {Xu}}]{arXiv:2202.12151}%
	\BibitemOpen
	\bibfield  {author} {\bibinfo {author} {\bibfnamefont {C.~B.}\ \bibnamefont
			{Hua}}, \bibinfo {author} {\bibfnamefont {F.}~\bibnamefont {Xiao}}, \bibinfo
		{author} {\bibfnamefont {Z.~R.}\ \bibnamefont {Liu}}, \bibinfo {author}
		{\bibfnamefont {J.~H.}\ \bibnamefont {Sun}}, \bibinfo {author} {\bibfnamefont
			{J.~H.}\ \bibnamefont {Gao}}, \bibinfo {author} {\bibfnamefont {C.~Z.}\
			\bibnamefont {Chen}}, \bibinfo {author} {\bibfnamefont {Q.}~\bibnamefont
			{Tong}}, \bibinfo {author} {\bibfnamefont {B.}~\bibnamefont {Zhou}}, \ and\
		\bibinfo {author} {\bibfnamefont {D.~H.}\ \bibnamefont {Xu}},\ }\bibfield
	{title} {\enquote {\bibinfo {title} {Magnon corner states in twisted bilayer
				honeycomb magnets},}\ }\href {\doibase 10.48550/arXiv.2202.12151} {\bibfield
		{journal} {\bibinfo  {journal} {arXiv:2202.12151}\ ,\ \bibinfo {pages}
			{224203}} (\bibinfo {year} {2022})}\BibitemShut {NoStop}%
	\bibitem [{\citenamefont {Liu}\ \emph {et~al.}(2021{\natexlab{b}})\citenamefont
		{Liu}, \citenamefont {Xian}, \citenamefont {Mu}, \citenamefont {Zhao},
		\citenamefont {Liu}, \citenamefont {Rubio},\ and\ \citenamefont
		{Wang}}]{PhysRevLett.126.066401}%
	\BibitemOpen
	\bibfield  {author} {\bibinfo {author} {\bibfnamefont {B.}~\bibnamefont
			{Liu}}, \bibinfo {author} {\bibfnamefont {L.~D.}\ \bibnamefont {Xian}},
		\bibinfo {author} {\bibfnamefont {H.~M.}\ \bibnamefont {Mu}}, \bibinfo
		{author} {\bibfnamefont {G.}~\bibnamefont {Zhao}}, \bibinfo {author}
		{\bibfnamefont {Z.}~\bibnamefont {Liu}}, \bibinfo {author} {\bibfnamefont
			{A.}~\bibnamefont {Rubio}}, \ and\ \bibinfo {author} {\bibfnamefont {Z.~F.}\
			\bibnamefont {Wang}},\ }\bibfield  {title} {\enquote {\bibinfo {title}
			{Higher-order band topology in twisted moir\'e superlattice},}\ }\href
	{\doibase 10.1103/PhysRevLett.126.066401} {\bibfield  {journal} {\bibinfo
			{journal} {Phys. Rev. Lett.}\ }\textbf {\bibinfo {volume} {126}},\ \bibinfo
		{pages} {066401} (\bibinfo {year} {2021}{\natexlab{b}})}\BibitemShut
	{NoStop}%
	\bibitem [{\citenamefont {Slager}\ \emph {et~al.}(2015)\citenamefont {Slager},
		\citenamefont {Rademaker}, \citenamefont {Zaanen},\ and\ \citenamefont
		{Balents}}]{PhysRevB.92.085126}%
	\BibitemOpen
	\bibfield  {author} {\bibinfo {author} {\bibfnamefont {R.~J.}\ \bibnamefont
			{Slager}}, \bibinfo {author} {\bibfnamefont {L.}~\bibnamefont {Rademaker}},
		\bibinfo {author} {\bibfnamefont {J.}~\bibnamefont {Zaanen}}, \ and\ \bibinfo
		{author} {\bibfnamefont {L.}~\bibnamefont {Balents}},\ }\bibfield  {title}
	{\enquote {\bibinfo {title} {Impurity-bound states and green's function zeros
				as local signatures of topology},}\ }\href {\doibase
		10.1103/PhysRevB.92.085126} {\bibfield  {journal} {\bibinfo  {journal} {Phys.
				Rev. B}\ }\textbf {\bibinfo {volume} {92}},\ \bibinfo {pages} {085126}
		(\bibinfo {year} {2015})}\BibitemShut {NoStop}%
	\bibitem [{\citenamefont {Ghosh}\ \emph {et~al.}(2021)\citenamefont {Ghosh},
		\citenamefont {Nag},\ and\ \citenamefont {Saha}}]{PhysRevB.104.134508}%
	\BibitemOpen
	\bibfield  {author} {\bibinfo {author} {\bibfnamefont {A.~K.}\ \bibnamefont
			{Ghosh}}, \bibinfo {author} {\bibfnamefont {T.}~\bibnamefont {Nag}}, \ and\
		\bibinfo {author} {\bibfnamefont {A.}~\bibnamefont {Saha}},\ }\bibfield
	{title} {\enquote {\bibinfo {title} {Hierarchy of higher-order topological
				superconductors in three dimensions},}\ }\href {\doibase
		10.1103/PhysRevB.104.134508} {\bibfield  {journal} {\bibinfo  {journal}
			{Phys. Rev. B}\ }\textbf {\bibinfo {volume} {104}},\ \bibinfo {pages}
		{134508} (\bibinfo {year} {2021})}\BibitemShut {NoStop}%
	\bibitem [{\citenamefont {Imhof}\ \emph {et~al.}(2018)\citenamefont {Imhof},
		\citenamefont {Berger}, \citenamefont {Bayer}, \citenamefont {Brehm},
		\citenamefont {Molenkamp}, \citenamefont {Kiessling}, \citenamefont
		{Schindler}, \citenamefont {Lee}, \citenamefont {Greiter}, \citenamefont
		{Neupert} \emph {et~al.}}]{imhof2018topolectrical}%
	\BibitemOpen
	\bibfield  {author} {\bibinfo {author} {\bibfnamefont {S.}~\bibnamefont
			{Imhof}}, \bibinfo {author} {\bibfnamefont {C.}~\bibnamefont {Berger}},
		\bibinfo {author} {\bibfnamefont {F.}~\bibnamefont {Bayer}}, \bibinfo
		{author} {\bibfnamefont {J.}~\bibnamefont {Brehm}}, \bibinfo {author}
		{\bibfnamefont {L.~W.}\ \bibnamefont {Molenkamp}}, \bibinfo {author}
		{\bibfnamefont {T.}~\bibnamefont {Kiessling}}, \bibinfo {author}
		{\bibfnamefont {F.}~\bibnamefont {Schindler}}, \bibinfo {author}
		{\bibfnamefont {C.~H.}\ \bibnamefont {Lee}}, \bibinfo {author} {\bibfnamefont
			{M.}~\bibnamefont {Greiter}}, \bibinfo {author} {\bibfnamefont
			{T.}~\bibnamefont {Neupert}},  \emph {et~al.},\ }\bibfield  {title} {\enquote
		{\bibinfo {title} {Topolectrical-circuit realization of topological corner
				modes},}\ }\href {\doibase https://doi.org/10.1038/s41567-018-0246-1}
	{\bibfield  {journal} {\bibinfo  {journal} {Nature Physics}\ }\textbf
		{\bibinfo {volume} {14}},\ \bibinfo {pages} {925--929} (\bibinfo {year}
		{2018})}\BibitemShut {NoStop}%
	\bibitem [{\citenamefont {Serra-Garcia}\ \emph {et~al.}(2019)\citenamefont
		{Serra-Garcia}, \citenamefont {S\"usstrunk},\ and\ \citenamefont
		{Huber}}]{PhysRevB.99.020304}%
	\BibitemOpen
	\bibfield  {author} {\bibinfo {author} {\bibfnamefont {M.}~\bibnamefont
			{Serra-Garcia}}, \bibinfo {author} {\bibfnamefont {R.}~\bibnamefont
			{S\"usstrunk}}, \ and\ \bibinfo {author} {\bibfnamefont {S.~D.}\ \bibnamefont
			{Huber}},\ }\bibfield  {title} {\enquote {\bibinfo {title} {Observation of
				quadrupole transitions and edge mode topology in an lc circuit network},}\
	}\href {\doibase 10.1103/PhysRevB.99.020304} {\bibfield  {journal} {\bibinfo
			{journal} {Phys. Rev. B}\ }\textbf {\bibinfo {volume} {99}},\ \bibinfo
		{pages} {020304} (\bibinfo {year} {2019})}\BibitemShut {NoStop}%
	\bibitem [{\citenamefont {Zhang}\ \emph {et~al.}(2021)\citenamefont {Zhang},
		\citenamefont {Zou}, \citenamefont {Pei}, \citenamefont {He}, \citenamefont
		{Bao}, \citenamefont {Sun},\ and\ \citenamefont
		{Zhang}}]{PhysRevLett.126.146802}%
	\BibitemOpen
	\bibfield  {author} {\bibinfo {author} {\bibfnamefont {W.~X.}\ \bibnamefont
			{Zhang}}, \bibinfo {author} {\bibfnamefont {D.~Y.}\ \bibnamefont {Zou}},
		\bibinfo {author} {\bibfnamefont {Q.~S.}\ \bibnamefont {Pei}}, \bibinfo
		{author} {\bibfnamefont {W.~J.}\ \bibnamefont {He}}, \bibinfo {author}
		{\bibfnamefont {J.~C.}\ \bibnamefont {Bao}}, \bibinfo {author} {\bibfnamefont
			{H.~J.}\ \bibnamefont {Sun}}, \ and\ \bibinfo {author} {\bibfnamefont
			{X.~D.}\ \bibnamefont {Zhang}},\ }\bibfield  {title} {\enquote {\bibinfo
			{title} {Experimental observation of higher-order topological anderson
				insulators},}\ }\href {\doibase 10.1103/PhysRevLett.126.146802} {\bibfield
		{journal} {\bibinfo  {journal} {Phys. Rev. Lett.}\ }\textbf {\bibinfo
			{volume} {126}},\ \bibinfo {pages} {146802} (\bibinfo {year}
		{2021})}\BibitemShut {NoStop}%
	\bibitem [{\citenamefont {Xue}\ \emph {et~al.}(2020)\citenamefont {Xue},
		\citenamefont {Ge}, \citenamefont {Sun}, \citenamefont {Wang}, \citenamefont
		{Jia}, \citenamefont {Guan}, \citenamefont {Yuan}, \citenamefont {Chong},\
		and\ \citenamefont {Zhang}}]{xue2020observation}%
	\BibitemOpen
	\bibfield  {author} {\bibinfo {author} {\bibfnamefont {H.~R.}\ \bibnamefont
			{Xue}}, \bibinfo {author} {\bibfnamefont {Y.}~\bibnamefont {Ge}}, \bibinfo
		{author} {\bibfnamefont {H.~X.}\ \bibnamefont {Sun}}, \bibinfo {author}
		{\bibfnamefont {Q.}~\bibnamefont {Wang}}, \bibinfo {author} {\bibfnamefont
			{D.}~\bibnamefont {Jia}}, \bibinfo {author} {\bibfnamefont {Y.~J.}\
			\bibnamefont {Guan}}, \bibinfo {author} {\bibfnamefont {S.~Q.}\ \bibnamefont
			{Yuan}}, \bibinfo {author} {\bibfnamefont {Y.~D.}\ \bibnamefont {Chong}}, \
		and\ \bibinfo {author} {\bibfnamefont {B.}~\bibnamefont {Zhang}},\ }\bibfield
	{title} {\enquote {\bibinfo {title} {Observation of an acoustic octupole
				topological insulator},}\ }\href {https://doi.org/10.1038/s41467-020-16350-1}
	{\bibfield  {journal} {\bibinfo  {journal} {Nature Communications}\ }\textbf
		{\bibinfo {volume} {11}},\ \bibinfo {pages} {2442} (\bibinfo {year}
		{2020})}\BibitemShut {NoStop}%
	\bibitem [{\citenamefont {Gao}\ \emph {et~al.}(2021)\citenamefont {Gao},
		\citenamefont {Xue}, \citenamefont {Gu}, \citenamefont {Liu}, \citenamefont
		{Zhu},\ and\ \citenamefont {Zhang}}]{Gao_2021}%
	\BibitemOpen
	\bibfield  {author} {\bibinfo {author} {\bibfnamefont {H.}~\bibnamefont
			{Gao}}, \bibinfo {author} {\bibfnamefont {H.~R.}\ \bibnamefont {Xue}},
		\bibinfo {author} {\bibfnamefont {Z.~M.}\ \bibnamefont {Gu}}, \bibinfo
		{author} {\bibfnamefont {T.}~\bibnamefont {Liu}}, \bibinfo {author}
		{\bibfnamefont {J.}~\bibnamefont {Zhu}}, \ and\ \bibinfo {author}
		{\bibfnamefont {B.~L.}\ \bibnamefont {Zhang}},\ }\bibfield  {title} {\enquote
		{\bibinfo {title} {Non-hermitian route to higher-order topology in an
				acoustic crystal},}\ }\href {\doibase 10.1038/s41467-021-22223-y} {\bibfield
		{journal} {\bibinfo  {journal} {Nature Communications}\ }\textbf {\bibinfo
			{volume} {12}} (\bibinfo {year} {2019}),\
		10.1038/s41467-021-22223-y}\BibitemShut {NoStop}%
	\bibitem [{\citenamefont {Mittal}\ \emph {et~al.}(2019)\citenamefont {Mittal},
		\citenamefont {Orre}, \citenamefont {Zhu}, \citenamefont {Gorlach},
		\citenamefont {Poddubny},\ and\ \citenamefont {Hafezi}}]{Mittal2019}%
	\BibitemOpen
	\bibfield  {author} {\bibinfo {author} {\bibfnamefont {S.}~\bibnamefont
			{Mittal}}, \bibinfo {author} {\bibfnamefont {V.~Vikram}\ \bibnamefont
			{Orre}}, \bibinfo {author} {\bibfnamefont {G.}~\bibnamefont {Zhu}}, \bibinfo
		{author} {\bibfnamefont {M.~A.}\ \bibnamefont {Gorlach}}, \bibinfo {author}
		{\bibfnamefont {A.}~\bibnamefont {Poddubny}}, \ and\ \bibinfo {author}
		{\bibfnamefont {M.}~\bibnamefont {Hafezi}},\ }\bibfield  {title} {\enquote
		{\bibinfo {title} {Photonic quadrupole topological phases},}\ }\href
	{\doibase 10.1038/s41566-019-0452-0} {\bibfield  {journal} {\bibinfo
			{journal} {Nat. Photon.}\ }\textbf {\bibinfo {volume} {13}},\ \bibinfo
		{pages} {692} (\bibinfo {year} {2019})}\BibitemShut {NoStop}%
	\bibitem [{\citenamefont {Hassan}\ \emph {et~al.}(2019)\citenamefont {Hassan},
		\citenamefont {Kunst}, \citenamefont {Moritz}, \citenamefont {Andler},
		\citenamefont {Bergholtz},\ and\ \citenamefont {Bourennane}}]{el2019corner}%
	\BibitemOpen
	\bibfield  {author} {\bibinfo {author} {\bibfnamefont {A.~E.}\ \bibnamefont
			{Hassan}}, \bibinfo {author} {\bibfnamefont {F.~K.}\ \bibnamefont {Kunst}},
		\bibinfo {author} {\bibfnamefont {A.}~\bibnamefont {Moritz}}, \bibinfo
		{author} {\bibfnamefont {G.}~\bibnamefont {Andler}}, \bibinfo {author}
		{\bibfnamefont {E.~J.}\ \bibnamefont {Bergholtz}}, \ and\ \bibinfo {author}
		{\bibfnamefont {M.}~\bibnamefont {Bourennane}},\ }\bibfield  {title}
	{\enquote {\bibinfo {title} {Corner states of light in photonic
				waveguides},}\ }\href {\doibase https://doi.org/10.1038/s41566-019-0519-y}
	{\bibfield  {journal} {\bibinfo  {journal} {Nature Photonics}\ }\textbf
		{\bibinfo {volume} {13}},\ \bibinfo {pages} {697--700} (\bibinfo {year}
		{2019})}\BibitemShut {NoStop}%
	\bibitem [{\citenamefont {Li}\ \emph {et~al.}(2019)\citenamefont {Li},
		\citenamefont {Zhirihin}, \citenamefont {Gorlach}, \citenamefont {Ni},
		\citenamefont {Filonov}, \citenamefont {Slobozhanyuk}, \citenamefont
		{Al{\`{u}}},\ and\ \citenamefont {Khanikaev}}]{Li2019}%
	\BibitemOpen
	\bibfield  {author} {\bibinfo {author} {\bibfnamefont {M.}~\bibnamefont
			{Li}}, \bibinfo {author} {\bibfnamefont {D.}~\bibnamefont {Zhirihin}},
		\bibinfo {author} {\bibfnamefont {M.}~\bibnamefont {Gorlach}}, \bibinfo
		{author} {\bibfnamefont {X.}~\bibnamefont {Ni}}, \bibinfo {author}
		{\bibfnamefont {D.}~\bibnamefont {Filonov}}, \bibinfo {author} {\bibfnamefont
			{A.}~\bibnamefont {Slobozhanyuk}}, \bibinfo {author} {\bibfnamefont
			{A.}~\bibnamefont {Al{\`{u}}}}, \ and\ \bibinfo {author} {\bibfnamefont
			{A.~B.}\ \bibnamefont {Khanikaev}},\ }\bibfield  {title} {\enquote {\bibinfo
			{title} {Higher-order topological states in photonic kagome crystals with
				long-range interactions},}\ }\href {\doibase 10.1038/s41566-019-0561-9}
	{\bibfield  {journal} {\bibinfo  {journal} {Nat. Photon.}\ }\textbf {\bibinfo
			{volume} {14}},\ \bibinfo {pages} {89} (\bibinfo {year} {2019})}\BibitemShut
	{NoStop}%
	\bibitem [{\citenamefont {Wang}\ and\ \citenamefont
		{Wang}(2020)}]{PhysRevResearch.2.033521}%
	\BibitemOpen
	\bibfield  {author} {\bibinfo {author} {\bibfnamefont {C.}~\bibnamefont
			{Wang}}\ and\ \bibinfo {author} {\bibfnamefont {X.~R.}\ \bibnamefont
			{Wang}},\ }\bibfield  {title} {\enquote {\bibinfo {title} {Disorder-induced
				quantum phase transitions in three-dimensional second-order topological
				insulators},}\ }\href {\doibase 10.1103/PhysRevResearch.2.033521} {\bibfield
		{journal} {\bibinfo  {journal} {Phys. Rev. Research}\ }\textbf {\bibinfo
			{volume} {2}},\ \bibinfo {pages} {033521} (\bibinfo {year}
		{2020})}\BibitemShut {NoStop}%
	\bibitem [{\citenamefont {Li}\ \emph {et~al.}(2020)\citenamefont {Li},
		\citenamefont {Fu}, \citenamefont {Hu}, \citenamefont {Li},\ and\
		\citenamefont {Shen}}]{PhysRevLett.125.166801}%
	\BibitemOpen
	\bibfield  {author} {\bibinfo {author} {\bibfnamefont {C.~A.}\ \bibnamefont
			{Li}}, \bibinfo {author} {\bibfnamefont {B.}~\bibnamefont {Fu}}, \bibinfo
		{author} {\bibfnamefont {Z.~A.}\ \bibnamefont {Hu}}, \bibinfo {author}
		{\bibfnamefont {J.}~\bibnamefont {Li}}, \ and\ \bibinfo {author}
		{\bibfnamefont {S.~Q.}\ \bibnamefont {Shen}},\ }\bibfield  {title} {\enquote
		{\bibinfo {title} {Topological phase transitions in disordered electric
				quadrupole insulators},}\ }\href {\doibase 10.1103/PhysRevLett.125.166801}
	{\bibfield  {journal} {\bibinfo  {journal} {Phys. Rev. Lett.}\ }\textbf
		{\bibinfo {volume} {125}},\ \bibinfo {pages} {166801} (\bibinfo {year}
		{2020})}\BibitemShut {NoStop}%
	\bibitem [{\citenamefont {Yang}\ \emph {et~al.}(2021)\citenamefont {Yang},
		\citenamefont {Li}, \citenamefont {Duan},\ and\ \citenamefont
		{Xu}}]{PhysRevB.103.085408}%
	\BibitemOpen
	\bibfield  {author} {\bibinfo {author} {\bibfnamefont {Y.~B.}\ \bibnamefont
			{Yang}}, \bibinfo {author} {\bibfnamefont {K.}~\bibnamefont {Li}}, \bibinfo
		{author} {\bibfnamefont {L.~M.}\ \bibnamefont {Duan}}, \ and\ \bibinfo
		{author} {\bibfnamefont {Y.}~\bibnamefont {Xu}},\ }\bibfield  {title}
	{\enquote {\bibinfo {title} {Higher-order topological anderson insulators},}\
	}\href {\doibase 10.1103/PhysRevB.103.085408} {\bibfield  {journal} {\bibinfo
			{journal} {Phys. Rev. B}\ }\textbf {\bibinfo {volume} {103}},\ \bibinfo
		{pages} {085408} (\bibinfo {year} {2021})}\BibitemShut {NoStop}%
	\bibitem [{\citenamefont {Spurrier}\ and\ \citenamefont
		{Cooper}(2020)}]{PhysRevResearch.2.033071}%
	\BibitemOpen
	\bibfield  {author} {\bibinfo {author} {\bibfnamefont {S.}~\bibnamefont
			{Spurrier}}\ and\ \bibinfo {author} {\bibfnamefont {N.~R.}\ \bibnamefont
			{Cooper}},\ }\bibfield  {title} {\enquote {\bibinfo {title} {Kane-mele with a
				twist: Quasicrystalline higher-order topological insulators with fractional
				mass kinks},}\ }\href {\doibase 10.1103/PhysRevResearch.2.033071} {\bibfield
		{journal} {\bibinfo  {journal} {Phys. Rev. Research}\ }\textbf {\bibinfo
			{volume} {2}},\ \bibinfo {pages} {033071} (\bibinfo {year}
		{2020})}\BibitemShut {NoStop}%
	\bibitem [{\citenamefont {Hua}\ \emph {et~al.}(2020)\citenamefont {Hua},
		\citenamefont {Chen}, \citenamefont {Zhou},\ and\ \citenamefont
		{Xu}}]{PhysRevB.102.241102}%
	\BibitemOpen
	\bibfield  {author} {\bibinfo {author} {\bibfnamefont {C.~B.}\ \bibnamefont
			{Hua}}, \bibinfo {author} {\bibfnamefont {R.}~\bibnamefont {Chen}}, \bibinfo
		{author} {\bibfnamefont {B.}~\bibnamefont {Zhou}}, \ and\ \bibinfo {author}
		{\bibfnamefont {D.~H.}\ \bibnamefont {Xu}},\ }\bibfield  {title} {\enquote
		{\bibinfo {title} {Higher-order topological insulator in a dodecagonal
				quasicrystal},}\ }\href {\doibase 10.1103/PhysRevB.102.241102} {\bibfield
		{journal} {\bibinfo  {journal} {Phys. Rev. B}\ }\textbf {\bibinfo {volume}
			{102}},\ \bibinfo {pages} {241102} (\bibinfo {year} {2020})}\BibitemShut
	{NoStop}%
	\bibitem [{\citenamefont {Agarwala}\ \emph {et~al.}(2020)\citenamefont
		{Agarwala}, \citenamefont {Juri\ifmmode \check{c}\else
			\v{c}\fi{}i\ifmmode~\acute{c}\else \'{c}\fi{}},\ and\ \citenamefont
		{Roy}}]{PhysRevResearch.2.012067}%
	\BibitemOpen
	\bibfield  {author} {\bibinfo {author} {\bibfnamefont {A.}~\bibnamefont
			{Agarwala}}, \bibinfo {author} {\bibfnamefont {V.}~\bibnamefont {Juri\ifmmode
				\check{c}\else \v{c}\fi{}i\ifmmode~\acute{c}\else \'{c}\fi{}}}, \ and\
		\bibinfo {author} {\bibfnamefont {B.}~\bibnamefont {Roy}},\ }\bibfield
	{title} {\enquote {\bibinfo {title} {Higher-order topological insulators in
				amorphous solids},}\ }\href {\doibase 10.1103/PhysRevResearch.2.012067}
	{\bibfield  {journal} {\bibinfo  {journal} {Phys. Rev. Research}\ }\textbf
		{\bibinfo {volume} {2}},\ \bibinfo {pages} {012067} (\bibinfo {year}
		{2020})}\BibitemShut {NoStop}%
	\bibitem [{\citenamefont {Wang}\ \emph {et~al.}(2021)\citenamefont {Wang},
		\citenamefont {Yang}, \citenamefont {Dai},\ and\ \citenamefont
		{Xu}}]{PhysRevLett.126.206404}%
	\BibitemOpen
	\bibfield  {author} {\bibinfo {author} {\bibfnamefont {J.~H.}\ \bibnamefont
			{Wang}}, \bibinfo {author} {\bibfnamefont {Y.~B.}\ \bibnamefont {Yang}},
		\bibinfo {author} {\bibfnamefont {N.}~\bibnamefont {Dai}}, \ and\ \bibinfo
		{author} {\bibfnamefont {Y.}~\bibnamefont {Xu}},\ }\bibfield  {title}
	{\enquote {\bibinfo {title} {Structural-disorder-induced second-order
				topological insulators in three dimensions},}\ }\href {\doibase
		10.1103/PhysRevLett.126.206404} {\bibfield  {journal} {\bibinfo  {journal}
			{Phys. Rev. Lett.}\ }\textbf {\bibinfo {volume} {126}},\ \bibinfo {pages}
		{206404} (\bibinfo {year} {2021})}\BibitemShut {NoStop}%
	\bibitem [{\citenamefont {Peng}\ \emph {et~al.}(2020)\citenamefont {Peng},
		\citenamefont {He},\ and\ \citenamefont {Lu}}]{PhysRevB.102.045110}%
	\BibitemOpen
	\bibfield  {author} {\bibinfo {author} {\bibfnamefont {C.}~\bibnamefont
			{Peng}}, \bibinfo {author} {\bibfnamefont {R.~Q.}\ \bibnamefont {He}}, \ and\
		\bibinfo {author} {\bibfnamefont {Z.~Y.}\ \bibnamefont {Lu}},\ }\bibfield
	{title} {\enquote {\bibinfo {title} {Correlation effects in quadrupole
				insulators: A quantum monte carlo study},}\ }\href {\doibase
		10.1103/PhysRevB.102.045110} {\bibfield  {journal} {\bibinfo  {journal}
			{Phys. Rev. B}\ }\textbf {\bibinfo {volume} {102}},\ \bibinfo {pages}
		{045110} (\bibinfo {year} {2020})}\BibitemShut {NoStop}%
	\bibitem [{\citenamefont {Kudo}\ \emph {et~al.}(2019)\citenamefont {Kudo},
		\citenamefont {Yoshida},\ and\ \citenamefont
		{Hatsugai}}]{PhysRevLett.123.196402}%
	\BibitemOpen
	\bibfield  {author} {\bibinfo {author} {\bibfnamefont {K.}~\bibnamefont
			{Kudo}}, \bibinfo {author} {\bibfnamefont {T.}~\bibnamefont {Yoshida}}, \
		and\ \bibinfo {author} {\bibfnamefont {Y.}~\bibnamefont {Hatsugai}},\
	}\bibfield  {title} {\enquote {\bibinfo {title} {Higher-order topological
				mott insulators},}\ }\href {\doibase 10.1103/PhysRevLett.123.196402}
	{\bibfield  {journal} {\bibinfo  {journal} {Phys. Rev. Lett.}\ }\textbf
		{\bibinfo {volume} {123}},\ \bibinfo {pages} {196402} (\bibinfo {year}
		{2019})}\BibitemShut {NoStop}%
	\bibitem [{\citenamefont {Zhao}\ \emph {et~al.}(2021)\citenamefont {Zhao},
		\citenamefont {Qiang}, \citenamefont {Lu},\ and\ \citenamefont
		{Xie}}]{PhysRevLett.127.176601}%
	\BibitemOpen
	\bibfield  {author} {\bibinfo {author} {\bibfnamefont {P.~L.}\ \bibnamefont
			{Zhao}}, \bibinfo {author} {\bibfnamefont {X.~B.}\ \bibnamefont {Qiang}},
		\bibinfo {author} {\bibfnamefont {H.~Z.}\ \bibnamefont {Lu}}, \ and\ \bibinfo
		{author} {\bibfnamefont {X.~C.}\ \bibnamefont {Xie}},\ }\bibfield  {title}
	{\enquote {\bibinfo {title} {Coulomb instabilities of a three-dimensional
				higher-order topological insulator},}\ }\href {\doibase
		10.1103/PhysRevLett.127.176601} {\bibfield  {journal} {\bibinfo  {journal}
			{Phys. Rev. Lett.}\ }\textbf {\bibinfo {volume} {127}},\ \bibinfo {pages}
		{176601} (\bibinfo {year} {2021})}\BibitemShut {NoStop}%
	\bibitem [{\citenamefont {Liu}\ \emph {et~al.}(2019)\citenamefont {Liu},
		\citenamefont {Zhang}, \citenamefont {Ai}, \citenamefont {Gong},
		\citenamefont {Kawabata}, \citenamefont {Ueda},\ and\ \citenamefont
		{Nori}}]{PhysRevLett.122.076801}%
	\BibitemOpen
	\bibfield  {author} {\bibinfo {author} {\bibfnamefont {T.}~\bibnamefont
			{Liu}}, \bibinfo {author} {\bibfnamefont {Y.~R.}\ \bibnamefont {Zhang}},
		\bibinfo {author} {\bibfnamefont {Q.}~\bibnamefont {Ai}}, \bibinfo {author}
		{\bibfnamefont {Z.~P.}\ \bibnamefont {Gong}}, \bibinfo {author}
		{\bibfnamefont {K.}~\bibnamefont {Kawabata}}, \bibinfo {author}
		{\bibfnamefont {M.}~\bibnamefont {Ueda}}, \ and\ \bibinfo {author}
		{\bibfnamefont {F.}~\bibnamefont {Nori}},\ }\bibfield  {title} {\enquote
		{\bibinfo {title} {Second-order topological phases in non-hermitian
				systems},}\ }\href {\doibase 10.1103/PhysRevLett.122.076801} {\bibfield
		{journal} {\bibinfo  {journal} {Phys. Rev. Lett.}\ }\textbf {\bibinfo
			{volume} {122}},\ \bibinfo {pages} {076801} (\bibinfo {year}
		{2019})}\BibitemShut {NoStop}%
	\bibitem [{\citenamefont {Luo}\ and\ \citenamefont
		{Zhang}(2019)}]{PhysRevLett.123.073601}%
	\BibitemOpen
	\bibfield  {author} {\bibinfo {author} {\bibfnamefont {X.~W.}\ \bibnamefont
			{Luo}}\ and\ \bibinfo {author} {\bibfnamefont {C.~W.}\ \bibnamefont
			{Zhang}},\ }\bibfield  {title} {\enquote {\bibinfo {title} {Higher-order
				topological corner states induced by gain and loss},}\ }\href {\doibase
		10.1103/PhysRevLett.123.073601} {\bibfield  {journal} {\bibinfo  {journal}
			{Phys. Rev. Lett.}\ }\textbf {\bibinfo {volume} {123}},\ \bibinfo {pages}
		{073601} (\bibinfo {year} {2019})}\BibitemShut {NoStop}%
	\bibitem [{\citenamefont {Liu}\ \emph {et~al.}(2021{\natexlab{c}})\citenamefont
		{Liu}, \citenamefont {Zhou}, \citenamefont {Wu}, \citenamefont {Zhang},\ and\
		\citenamefont {Jiang}}]{PhysRevB.103.224203}%
	\BibitemOpen
	\bibfield  {author} {\bibinfo {author} {\bibfnamefont {H.}~\bibnamefont
			{Liu}}, \bibinfo {author} {\bibfnamefont {J.~K.}\ \bibnamefont {Zhou}},
		\bibinfo {author} {\bibfnamefont {B.~L.}\ \bibnamefont {Wu}}, \bibinfo
		{author} {\bibfnamefont {Z.~Q.}\ \bibnamefont {Zhang}}, \ and\ \bibinfo
		{author} {\bibfnamefont {H.}~\bibnamefont {Jiang}},\ }\bibfield  {title}
	{\enquote {\bibinfo {title} {Real-space topological invariant and
				higher-order topological anderson insulator in two-dimensional non-hermitian
				systems},}\ }\href {\doibase 10.1103/PhysRevB.103.224203} {\bibfield
		{journal} {\bibinfo  {journal} {Phys. Rev. B}\ }\textbf {\bibinfo {volume}
			{103}},\ \bibinfo {pages} {224203} (\bibinfo {year}
		{2021}{\natexlab{c}})}\BibitemShut {NoStop}%
	\bibitem [{\citenamefont {Hu}\ \emph {et~al.}(2020)\citenamefont {Hu},
		\citenamefont {Huang}, \citenamefont {Zhao},\ and\ \citenamefont
		{Liu}}]{PhysRevLett.124.057001}%
	\BibitemOpen
	\bibfield  {author} {\bibinfo {author} {\bibfnamefont {H.~P.}\ \bibnamefont
			{Hu}}, \bibinfo {author} {\bibfnamefont {B.}~\bibnamefont {Huang}}, \bibinfo
		{author} {\bibfnamefont {E.}~\bibnamefont {Zhao}}, \ and\ \bibinfo {author}
		{\bibfnamefont {W.~V.}\ \bibnamefont {Liu}},\ }\bibfield  {title} {\enquote
		{\bibinfo {title} {Dynamical singularities of floquet higher-order
				topological insulators},}\ }\href {\doibase 10.1103/PhysRevLett.124.057001}
	{\bibfield  {journal} {\bibinfo  {journal} {Phys. Rev. Lett.}\ }\textbf
		{\bibinfo {volume} {124}},\ \bibinfo {pages} {057001} (\bibinfo {year}
		{2020})}\BibitemShut {NoStop}%
	\bibitem [{\citenamefont {Huang}\ and\ \citenamefont
		{Liu}(2020)}]{PhysRevLett.124.216601}%
	\BibitemOpen
	\bibfield  {author} {\bibinfo {author} {\bibfnamefont {B.}~\bibnamefont
			{Huang}}\ and\ \bibinfo {author} {\bibfnamefont {W.~V.}\ \bibnamefont
			{Liu}},\ }\bibfield  {title} {\enquote {\bibinfo {title} {Floquet
				higher-order topological insulators with anomalous dynamical polarization},}\
	}\href {\doibase 10.1103/PhysRevLett.124.216601} {\bibfield  {journal}
		{\bibinfo  {journal} {Phys. Rev. Lett.}\ }\textbf {\bibinfo {volume} {124}},\
		\bibinfo {pages} {216601} (\bibinfo {year} {2020})}\BibitemShut {NoStop}%
	\bibitem [{\citenamefont {Pan}\ and\ \citenamefont
		{Zhou}(2020)}]{PhysRevB.102.094305}%
	\BibitemOpen
	\bibfield  {author} {\bibinfo {author} {\bibfnamefont {J.}~\bibnamefont
			{Pan}}\ and\ \bibinfo {author} {\bibfnamefont {L.}~\bibnamefont {Zhou}},\
	}\bibfield  {title} {\enquote {\bibinfo {title} {Non-{H}ermitian {F}loquet
				second order topological insulators in periodically quenched lattices},}\
	}\href {\doibase 10.1103/PhysRevB.102.094305} {\bibfield  {journal} {\bibinfo
			{journal} {Phys. Rev. B}\ }\textbf {\bibinfo {volume} {102}},\ \bibinfo
		{pages} {094305} (\bibinfo {year} {2020})}\BibitemShut {NoStop}%
	\bibitem [{\citenamefont {Ghosh}\ \emph {et~al.}(2020)\citenamefont {Ghosh},
		\citenamefont {Paul},\ and\ \citenamefont {Saha}}]{PhysRevB.101.235403}%
	\BibitemOpen
	\bibfield  {author} {\bibinfo {author} {\bibfnamefont {A.~K.}\ \bibnamefont
			{Ghosh}}, \bibinfo {author} {\bibfnamefont {G.~C.}\ \bibnamefont {Paul}}, \
		and\ \bibinfo {author} {\bibfnamefont {A.}~\bibnamefont {Saha}},\ }\bibfield
	{title} {\enquote {\bibinfo {title} {Higher order topological insulator via
				periodic driving},}\ }\href {\doibase 10.1103/PhysRevB.101.235403} {\bibfield
		{journal} {\bibinfo  {journal} {Phys. Rev. B}\ }\textbf {\bibinfo {volume}
			{101}},\ \bibinfo {pages} {235403} (\bibinfo {year} {2020})}\BibitemShut
	{NoStop}%
	\bibitem [{\citenamefont {Ghosh}\ \emph
		{et~al.}(2022{\natexlab{a}})\citenamefont {Ghosh}, \citenamefont {Nag},\ and\
		\citenamefont {Saha}}]{PhysRevB.105.115418}%
	\BibitemOpen
	\bibfield  {author} {\bibinfo {author} {\bibfnamefont {A.~K.}\ \bibnamefont
			{Ghosh}}, \bibinfo {author} {\bibfnamefont {T.}~\bibnamefont {Nag}}, \ and\
		\bibinfo {author} {\bibfnamefont {A.}~\bibnamefont {Saha}},\ }\bibfield
	{title} {\enquote {\bibinfo {title} {Systematic generation of the cascade of
				anomalous dynamical first- and higher-order modes in floquet topological
				insulators},}\ }\href {\doibase 10.1103/PhysRevB.105.115418} {\bibfield
		{journal} {\bibinfo  {journal} {Phys. Rev. B}\ }\textbf {\bibinfo {volume}
			{105}},\ \bibinfo {pages} {115418} (\bibinfo {year}
		{2022}{\natexlab{a}})}\BibitemShut {NoStop}%
	\bibitem [{\citenamefont {Nag}\ \emph {et~al.}(2021)\citenamefont {Nag},
		\citenamefont {Juri\ifmmode \check{c}\else \v{c}\fi{}i\ifmmode~\acute{c}\else
			\'{c}\fi{}},\ and\ \citenamefont {Roy}}]{PhysRevB.103.115308}%
	\BibitemOpen
	\bibfield  {author} {\bibinfo {author} {\bibfnamefont {T.}~\bibnamefont
			{Nag}}, \bibinfo {author} {\bibfnamefont {V.}~\bibnamefont {Juri\ifmmode
				\check{c}\else \v{c}\fi{}i\ifmmode~\acute{c}\else \'{c}\fi{}}}, \ and\
		\bibinfo {author} {\bibfnamefont {B.}~\bibnamefont {Roy}},\ }\bibfield
	{title} {\enquote {\bibinfo {title} {Hierarchy of higher-order floquet
				topological phases in three dimensions},}\ }\href {\doibase
		10.1103/PhysRevB.103.115308} {\bibfield  {journal} {\bibinfo  {journal}
			{Phys. Rev. B}\ }\textbf {\bibinfo {volume} {103}},\ \bibinfo {pages}
		{115308} (\bibinfo {year} {2021})}\BibitemShut {NoStop}%
	\bibitem [{\citenamefont {Ghosh}\ \emph
		{et~al.}(2022{\natexlab{b}})\citenamefont {Ghosh}, \citenamefont {Nag},\ and\
		\citenamefont {Saha}}]{PhysRevB.105.155406}%
	\BibitemOpen
	\bibfield  {author} {\bibinfo {author} {\bibfnamefont {A.~K.}\ \bibnamefont
			{Ghosh}}, \bibinfo {author} {\bibfnamefont {T.}~\bibnamefont {Nag}}, \ and\
		\bibinfo {author} {\bibfnamefont {A.}~\bibnamefont {Saha}},\ }\bibfield
	{title} {\enquote {\bibinfo {title} {Dynamical construction of quadrupolar
				and octupolar topological superconductors},}\ }\href {\doibase
		10.1103/PhysRevB.105.155406} {\bibfield  {journal} {\bibinfo  {journal}
			{Phys. Rev. B}\ }\textbf {\bibinfo {volume} {105}},\ \bibinfo {pages}
		{155406} (\bibinfo {year} {2022}{\natexlab{b}})}\BibitemShut {NoStop}%
	\bibitem [{\citenamefont {Garate}(2013)}]{PhysRevLett.110.046402}%
	\BibitemOpen
	\bibfield  {author} {\bibinfo {author} {\bibfnamefont {I.}~\bibnamefont
			{Garate}},\ }\bibfield  {title} {\enquote {\bibinfo {title} {Phonon-induced
				topological transitions and crossovers in dirac materials},}\ }\href
	{\doibase 10.1103/PhysRevLett.110.046402} {\bibfield  {journal} {\bibinfo
			{journal} {Phys. Rev. Lett.}\ }\textbf {\bibinfo {volume} {110}},\ \bibinfo
		{pages} {046402} (\bibinfo {year} {2013})}\BibitemShut {NoStop}%
	\bibitem [{\citenamefont {M{\"o}ller}\ \emph {et~al.}(2017)\citenamefont
		{M{\"o}ller}, \citenamefont {Sawatzky}, \citenamefont {Franz},\ and\
		\citenamefont {Berciu}}]{2017Type}%
	\BibitemOpen
	\bibfield  {author} {\bibinfo {author} {\bibfnamefont {M.~M.}\ \bibnamefont
			{M{\"o}ller}}, \bibinfo {author} {\bibfnamefont {G.~A.}\ \bibnamefont
			{Sawatzky}}, \bibinfo {author} {\bibfnamefont {M.}~\bibnamefont {Franz}}, \
		and\ \bibinfo {author} {\bibfnamefont {M.}~\bibnamefont {Berciu}},\
	}\bibfield  {title} {\enquote {\bibinfo {title} {{Type-II} dirac semimetal
				stabilized by electron-phonon coupling},}\ }\href {\doibase
		10.1038/s41467-017-02442-y} {\bibfield  {journal} {\bibinfo  {journal} {Nat.
				Commun.}\ }\textbf {\bibinfo {volume} {8}},\ \bibinfo {pages} {2267}
		(\bibinfo {year} {2017})}\BibitemShut {NoStop}%
	\bibitem [{\citenamefont {Gonz\'alez-Cuadra}\ \emph {et~al.}(2018)\citenamefont
		{Gonz\'alez-Cuadra}, \citenamefont {Grzybowski}, \citenamefont {Dauphin},\
		and\ \citenamefont {Lewenstein}}]{PhysRevLett.121.090402}%
	\BibitemOpen
	\bibfield  {author} {\bibinfo {author} {\bibfnamefont {D.}~\bibnamefont
			{Gonz\'alez-Cuadra}}, \bibinfo {author} {\bibfnamefont {P.~R.}\ \bibnamefont
			{Grzybowski}}, \bibinfo {author} {\bibfnamefont {A.}~\bibnamefont {Dauphin}},
		\ and\ \bibinfo {author} {\bibfnamefont {M.}~\bibnamefont {Lewenstein}},\
	}\bibfield  {title} {\enquote {\bibinfo {title} {Strongly correlated bosons
				on a dynamical lattice},}\ }\href {\doibase 10.1103/PhysRevLett.121.090402}
	{\bibfield  {journal} {\bibinfo  {journal} {Phys. Rev. Lett.}\ }\textbf
		{\bibinfo {volume} {121}},\ \bibinfo {pages} {090402} (\bibinfo {year}
		{2018})}\BibitemShut {NoStop}%
	\bibitem [{\citenamefont {Antonius}\ and\ \citenamefont
		{Louie}(2016)}]{PhysRevLett.117.246401}%
	\BibitemOpen
	\bibfield  {author} {\bibinfo {author} {\bibfnamefont {G.}~\bibnamefont
			{Antonius}}\ and\ \bibinfo {author} {\bibfnamefont {S.~G.}\ \bibnamefont
			{Louie}},\ }\bibfield  {title} {\enquote {\bibinfo {title}
			{Temperature-induced topological phase transitions: Promoted versus
				suppressed nontrivial topology},}\ }\href {\doibase
		10.1103/PhysRevLett.117.246401} {\bibfield  {journal} {\bibinfo  {journal}
			{Phys. Rev. Lett.}\ }\textbf {\bibinfo {volume} {117}},\ \bibinfo {pages}
		{246401} (\bibinfo {year} {2016})}\BibitemShut {NoStop}%
	\bibitem [{\citenamefont {Cangemi}\ \emph {et~al.}(2019)\citenamefont
		{Cangemi}, \citenamefont {Mishchenko}, \citenamefont {Nagaosa}, \citenamefont
		{Cataudella},\ and\ \citenamefont {De~Filippis}}]{PhysRevLett.123.046401}%
	\BibitemOpen
	\bibfield  {author} {\bibinfo {author} {\bibfnamefont {L.~M.}\ \bibnamefont
			{Cangemi}}, \bibinfo {author} {\bibfnamefont {A.~S.}\ \bibnamefont
			{Mishchenko}}, \bibinfo {author} {\bibfnamefont {N.}~\bibnamefont {Nagaosa}},
		\bibinfo {author} {\bibfnamefont {V.}~\bibnamefont {Cataudella}}, \ and\
		\bibinfo {author} {\bibfnamefont {G.}~\bibnamefont {De~Filippis}},\
	}\bibfield  {title} {\enquote {\bibinfo {title} {Topological quantum
				transition driven by charge-phonon coupling in the haldane chern
				insulator},}\ }\href {\doibase 10.1103/PhysRevLett.123.046401} {\bibfield
		{journal} {\bibinfo  {journal} {Phys. Rev. Lett.}\ }\textbf {\bibinfo
			{volume} {123}},\ \bibinfo {pages} {046401} (\bibinfo {year}
		{2019})}\BibitemShut {NoStop}%
	\bibitem [{\citenamefont {Calvo}\ \emph {et~al.}(2018)\citenamefont {Calvo},
		\citenamefont {Luna}, \citenamefont {Dal~Lago},\ and\ \citenamefont
		{Foa~Torres}}]{PhysRevB.98.035423}%
	\BibitemOpen
	\bibfield  {author} {\bibinfo {author} {\bibfnamefont {H.~L.}\ \bibnamefont
			{Calvo}}, \bibinfo {author} {\bibfnamefont {J.~S.}\ \bibnamefont {Luna}},
		\bibinfo {author} {\bibfnamefont {V.}~\bibnamefont {Dal~Lago}}, \ and\
		\bibinfo {author} {\bibfnamefont {L.~E.~F.}\ \bibnamefont {Foa~Torres}},\
	}\bibfield  {title} {\enquote {\bibinfo {title} {Robust edge states induced
				by electron-phonon interaction in graphene nanoribbons},}\ }\href {\doibase
		10.1103/PhysRevB.98.035423} {\bibfield  {journal} {\bibinfo  {journal} {Phys.
				Rev. B}\ }\textbf {\bibinfo {volume} {98}},\ \bibinfo {pages} {035423}
		(\bibinfo {year} {2018})}\BibitemShut {NoStop}%
	\bibitem [{\citenamefont {Chaudhary}\ \emph {et~al.}(2020)\citenamefont
		{Chaudhary}, \citenamefont {Haim}, \citenamefont {Peng},\ and\ \citenamefont
		{Refael}}]{PhysRevResearch.2.043431}%
	\BibitemOpen
	\bibfield  {author} {\bibinfo {author} {\bibfnamefont {S.}~\bibnamefont
			{Chaudhary}}, \bibinfo {author} {\bibfnamefont {A.}~\bibnamefont {Haim}},
		\bibinfo {author} {\bibfnamefont {Y.}~\bibnamefont {Peng}}, \ and\ \bibinfo
		{author} {\bibfnamefont {G.}~\bibnamefont {Refael}},\ }\bibfield  {title}
	{\enquote {\bibinfo {title} {Phonon-induced floquet topological phases
				protected by space-time symmetries},}\ }\href {\doibase
		10.1103/PhysRevResearch.2.043431} {\bibfield  {journal} {\bibinfo  {journal}
			{Phys. Rev. Research}\ }\textbf {\bibinfo {volume} {2}},\ \bibinfo {pages}
		{043431} (\bibinfo {year} {2020})}\BibitemShut {NoStop}%
	\bibitem [{\citenamefont {Medina Due\~nas}\ \emph {et~al.}(2022)\citenamefont
		{Medina Due\~nas}, \citenamefont {Calvo},\ and\ \citenamefont {Foa T.and
			Foa~Torres}}]{PhysRevLett.128.066801}%
	\BibitemOpen
	\bibfield  {author} {\bibinfo {author} {\bibfnamefont {J.}~\bibnamefont
			{Medina Due\~nas}}, \bibinfo {author} {\bibfnamefont {H.~L.}\ \bibnamefont
			{Calvo}}, \ and\ \bibinfo {author} {\bibfnamefont {L.~E.~F.}\ \bibnamefont
			{Foa T.and Foa~Torres}},\ }\bibfield  {title} {\enquote {\bibinfo {title}
			{Copropagating edge states produced by the interaction between electrons and
				chiral phonons in two-dimensional materials},}\ }\href {\doibase
		10.1103/PhysRevLett.128.066801} {\bibfield  {journal} {\bibinfo  {journal}
			{Phys. Rev. Lett.}\ }\textbf {\bibinfo {volume} {128}},\ \bibinfo {pages}
		{066801} (\bibinfo {year} {2022})}\BibitemShut {NoStop}%
	\bibitem [{\citenamefont {Ren}\ \emph {et~al.}(2020)\citenamefont {Ren},
		\citenamefont {Qiao},\ and\ \citenamefont {Niu}}]{PhysRevLett.124.166804}%
	\BibitemOpen
	\bibfield  {author} {\bibinfo {author} {\bibfnamefont {Y.~F.}\ \bibnamefont
			{Ren}}, \bibinfo {author} {\bibfnamefont {Z.~H.}\ \bibnamefont {Qiao}}, \
		and\ \bibinfo {author} {\bibfnamefont {Q.}~\bibnamefont {Niu}},\ }\bibfield
	{title} {\enquote {\bibinfo {title} {Engineering corner states from
				two-dimensional topological insulators},}\ }\href {\doibase
		10.1103/PhysRevLett.124.166804} {\bibfield  {journal} {\bibinfo  {journal}
			{Phys. Rev. Lett.}\ }\textbf {\bibinfo {volume} {124}},\ \bibinfo {pages}
		{166804} (\bibinfo {year} {2020})}\BibitemShut {NoStop}%
	\bibitem [{\citenamefont {S\'en\'echal}\ \emph {et~al.}(2002)\citenamefont
		{S\'en\'echal}, \citenamefont {Perez},\ and\ \citenamefont
		{Plouffe}}]{PhysRevB.66.075129}%
	\BibitemOpen
	\bibfield  {author} {\bibinfo {author} {\bibfnamefont {D.}~\bibnamefont
			{S\'en\'echal}}, \bibinfo {author} {\bibfnamefont {D.}~\bibnamefont {Perez}},
		\ and\ \bibinfo {author} {\bibfnamefont {D.}~\bibnamefont {Plouffe}},\
	}\bibfield  {title} {\enquote {\bibinfo {title} {Cluster perturbation theory
				for hubbard models},}\ }\href {\doibase 10.1103/PhysRevB.66.075129}
	{\bibfield  {journal} {\bibinfo  {journal} {Phys. Rev. B}\ }\textbf {\bibinfo
			{volume} {66}},\ \bibinfo {pages} {075129} (\bibinfo {year}
		{2002})}\BibitemShut {NoStop}%
	\bibitem [{\citenamefont {Sbierski}\ and\ \citenamefont
		{Karrasch}(2018)}]{PhysRevB.98.165101}%
	\BibitemOpen
	\bibfield  {author} {\bibinfo {author} {\bibfnamefont {B.}~\bibnamefont
			{Sbierski}}\ and\ \bibinfo {author} {\bibfnamefont {C.}~\bibnamefont
			{Karrasch}},\ }\bibfield  {title} {\enquote {\bibinfo {title} {Topological
				invariants for the haldane phase of interacting su-schrieffer-heeger chains:
				Functional renormalization-group approach},}\ }\href {\doibase
		10.1103/PhysRevB.98.165101} {\bibfield  {journal} {\bibinfo  {journal} {Phys.
				Rev. B}\ }\textbf {\bibinfo {volume} {98}},\ \bibinfo {pages} {165101}
		(\bibinfo {year} {2018})}\BibitemShut {NoStop}%
	\bibitem [{\citenamefont {Ezawa}(2012)}]{PhysRevLett.109.055502}%
	\BibitemOpen
	\bibfield  {author} {\bibinfo {author} {\bibfnamefont {M.}~\bibnamefont
			{Ezawa}},\ }\bibfield  {title} {\enquote {\bibinfo {title} {Valley-polarized
				metals and quantum anomalous hall effect in silicene},}\ }\href {\doibase
		10.1103/PhysRevLett.109.055502} {\bibfield  {journal} {\bibinfo  {journal}
			{Phys. Rev. Lett.}\ }\textbf {\bibinfo {volume} {109}},\ \bibinfo {pages}
		{055502} (\bibinfo {year} {2012})}\BibitemShut {NoStop}%
	\bibitem [{\citenamefont {Marrazzo}\ \emph {et~al.}(2018)\citenamefont
		{Marrazzo}, \citenamefont {Gibertini}, \citenamefont {Campi}, \citenamefont
		{Mounet},\ and\ \citenamefont {Marzari}}]{PhysRevLett.120.117701}%
	\BibitemOpen
	\bibfield  {author} {\bibinfo {author} {\bibfnamefont {A.}~\bibnamefont
			{Marrazzo}}, \bibinfo {author} {\bibfnamefont {M.}~\bibnamefont {Gibertini}},
		\bibinfo {author} {\bibfnamefont {D.}~\bibnamefont {Campi}}, \bibinfo
		{author} {\bibfnamefont {N.}~\bibnamefont {Mounet}}, \ and\ \bibinfo {author}
		{\bibfnamefont {N.}~\bibnamefont {Marzari}},\ }\bibfield  {title} {\enquote
		{\bibinfo {title} {Prediction of a large-gap and switchable {K}ane-{M}ele
				quantum spin hall insulator},}\ }\href {\doibase
		10.1103/PhysRevLett.120.117701} {\bibfield  {journal} {\bibinfo  {journal}
			{Phys. Rev. Lett.}\ }\textbf {\bibinfo {volume} {120}},\ \bibinfo {pages}
		{117701} (\bibinfo {year} {2018})}\BibitemShut {NoStop}%
	\bibitem [{\citenamefont {Yan}\ \emph {et~al.}(2015)\citenamefont {Yan},
		\citenamefont {Gao}, \citenamefont {Stein},\ and\ \citenamefont
		{Coard}}]{PhysRevB.91.245403}%
	\BibitemOpen
	\bibfield  {author} {\bibinfo {author} {\bibfnamefont {J.-A.}\ \bibnamefont
			{Yan}}, \bibinfo {author} {\bibfnamefont {S.-P.}\ \bibnamefont {Gao}},
		\bibinfo {author} {\bibfnamefont {R.}~\bibnamefont {Stein}}, \ and\ \bibinfo
		{author} {\bibfnamefont {G.}~\bibnamefont {Coard}},\ }\bibfield  {title}
	{\enquote {\bibinfo {title} {Tuning the electronic structure of silicene and
				germanene by biaxial strain and electric field},}\ }\href {\doibase
		10.1103/PhysRevB.91.245403} {\bibfield  {journal} {\bibinfo  {journal} {Phys.
				Rev. B}\ }\textbf {\bibinfo {volume} {91}},\ \bibinfo {pages} {245403}
		(\bibinfo {year} {2015})}\BibitemShut {NoStop}%
	\bibitem [{\citenamefont {Hague}\ and\ \citenamefont
		{MacCormick}(2012)}]{Hague2012}%
	\BibitemOpen
	\bibfield  {author} {\bibinfo {author} {\bibfnamefont {J.~P.}\ \bibnamefont
			{Hague}}\ and\ \bibinfo {author} {\bibfnamefont {C.}~\bibnamefont
			{MacCormick}},\ }\bibfield  {title} {\enquote {\bibinfo {title} {Quantum
				simulation of electron{\textendash}phonon interactions in strongly deformable
				materials},}\ }\href {\doibase 10.1088/1367-2630/14/3/033019} {\bibfield
		{journal} {\bibinfo  {journal} {New J. Phys.}\ }\textbf {\bibinfo {volume}
			{14}},\ \bibinfo {pages} {033019} (\bibinfo {year} {2012})}\BibitemShut
	{NoStop}%
	\bibitem [{\citenamefont {Mendonca}\ and\ \citenamefont
		{Jachymski}(2022)}]{Mendonca}%
	\BibitemOpen
	\bibfield  {author} {\bibinfo {author} {\bibfnamefont {J.~P.}\ \bibnamefont
			{Mendonca}}\ and\ \bibinfo {author} {\bibfnamefont {K.}~\bibnamefont
			{Jachymski}},\ }\bibfield  {title} {\enquote {\bibinfo {title} {Quantum
				simulation of extended electron-phonon coupling models in a hybrid rydberg
				atom setup},}\ }\href@noop {} {\bibfield  {journal} {\bibinfo  {journal}
			{arXiv:2208.11473}\ } (\bibinfo {year} {2022})}\BibitemShut {NoStop}%
	\bibitem [{\citenamefont {Kn\"orzer}\ \emph {et~al.}(2022)\citenamefont
		{Kn\"orzer}, \citenamefont {Shi}, \citenamefont {Demler},\ and\ \citenamefont
		{Cirac}}]{PhysRevLett.128.120404}%
	\BibitemOpen
	\bibfield  {author} {\bibinfo {author} {\bibfnamefont {J.}~\bibnamefont
			{Kn\"orzer}}, \bibinfo {author} {\bibfnamefont {T.}~\bibnamefont {Shi}},
		\bibinfo {author} {\bibfnamefont {E.}~\bibnamefont {Demler}}, \ and\ \bibinfo
		{author} {\bibfnamefont {J.~I.}\ \bibnamefont {Cirac}},\ }\bibfield  {title}
	{\enquote {\bibinfo {title} {Spin-{H}olstein models in trapped-ion
				systems},}\ }\href {\doibase 10.1103/PhysRevLett.128.120404} {\bibfield
		{journal} {\bibinfo  {journal} {Phys. Rev. Lett.}\ }\textbf {\bibinfo
			{volume} {128}},\ \bibinfo {pages} {120404} (\bibinfo {year}
		{2022})}\BibitemShut {NoStop}%
\end{thebibliography}
\end{document}